\documentclass[4paper]{article}

\usepackage{cite,latexsym,amsmath,amssymb,tabularx,supertabular,bm,wrapfig,subfig,subfigure}

\newcommand{\tr}{{\rm tr}}               

\usepackage{pstricks,color}
\definecolor{lightgray}{gray}{.80}
\newgray{verylightgray}{.80}

\newif\ifpdf
\ifx\pdfoutput\undefined
     \pdffalse                           
     
     \usepackage{graphicx}
\else
     \pdfoutput=1
     \pdftrue
      
     \usepackage[backref,colorlinks=true]{hyperref}
     \usepackage[pdftex]{graphicx}    
\fi

\usepackage{rotating}

\usepackage{fancyheadings,calc,ifthen}
\newcounter{hours}\newcounter{minutes}
\newcommand{\printtime}{%
  \setcounter{hours}{\time/60}%
  \setcounter{minutes}{\time-\value{hours}*60}%
  \ifthenelse{\value{hours}<10}{0}{}\thehours:%
  \ifthenelse{\value{minutes}<10}{0}{}\theminutes}

\begin{document}
\hfill CERN-TH/2003-230
\vspace{1cm}

\noindent{\Large\bf Universal features of JIMWLK and BK evolution at small $x$}\\
  \vskip 0.5cm
  {\large Kari Rummukainen$,^{1,2}$ and Heribert Weigert${}^3$}\\

  {\small
  ${}^1$ CERN, Theory division, CH-1211 Geneva 23, Switzerland\\
  ${}^2$ Department of Physics, University of Oulu, P.O.Box 3000,
            FIN-90014 Oulu, Finland\\
  ${}^3$ Institut f\"ur theoretische Physik, 
          Universit\"at Regensburg, 93040 Regensburg,  Germany
   }
\vspace{.2cm}
\noindent\begin{center}
\begin{minipage}{.915\textwidth}
  {\small\sf In this paper we present the results of numerical studies
    of the JIMWLK and BK equations with a particular emphasis on the
    universal scaling properties and phase space structure involved.
    The results are valid for near zero impact parameter in DIS. We
    demonstrate IR safety due to the occurrence of a rapidity dependent
    saturation scale $Q_s(\tau)$.  Within the set of initial
    conditions chosen both JIMWLK and BK equations show remarkable
    agreement. We point out the crucial importance of running
    coupling corrections to obtain consistency in the UV.
    Despite the scale breaking induced by the running coupling we find
    that evolution drives correlators towards an asymptotic form with
    near scaling properties. We discuss asymptotic features of the
    evolution, such as the $\tau$- and $A$-dependence of $Q_s$ away
    from the initial condition.
  }
\end{minipage}
\end{center}
\vspace{1cm}

\section{Introduction}
\label{sec:introduction}

With the advent of modern colliders, from the Tevatron, HERA to RHIC
and planned experiments like LHC and EIC, the high energy asymptotics
of QCD has gained new prominence and importance. The question if QCD
effects there show universal features that might be calculated without
recourse to intrinsically non-perturbative methods capable of also
dealing with large $\alpha_s$ has gained new importance. In situations
in which the experimental probes see large gluon densities at small
$x$ the answer to this question has, maybe surprisingly, turned out to
be positive.  The progress has come from a method that resums the
``density'' induced corrections exactly while treating other effects,
such as the $x$ dependence as perturbative corrections. The
justification for this is that the density effects induce a generally
$x$ dependent correlation length $1/Q_s(x)$ which also sets the scale
for the coupling, thus giving us a small expansion parameter. There
are several properties of the interactions at small $x$ that allow us
to perform such a resummation. To explain them let us consider the
simplest example of a $\gamma^* A$ collision, deep inelastic scatting
(DIS) of virtual photons off nuclear targets, be it protons or larger
nuclei. Generalizations to diffraction, $p A$ and $A A$ scattering are
possible and share many of the key features.

\begin{figure}[htbp]
  \centering
  \includegraphics[width=6cm]{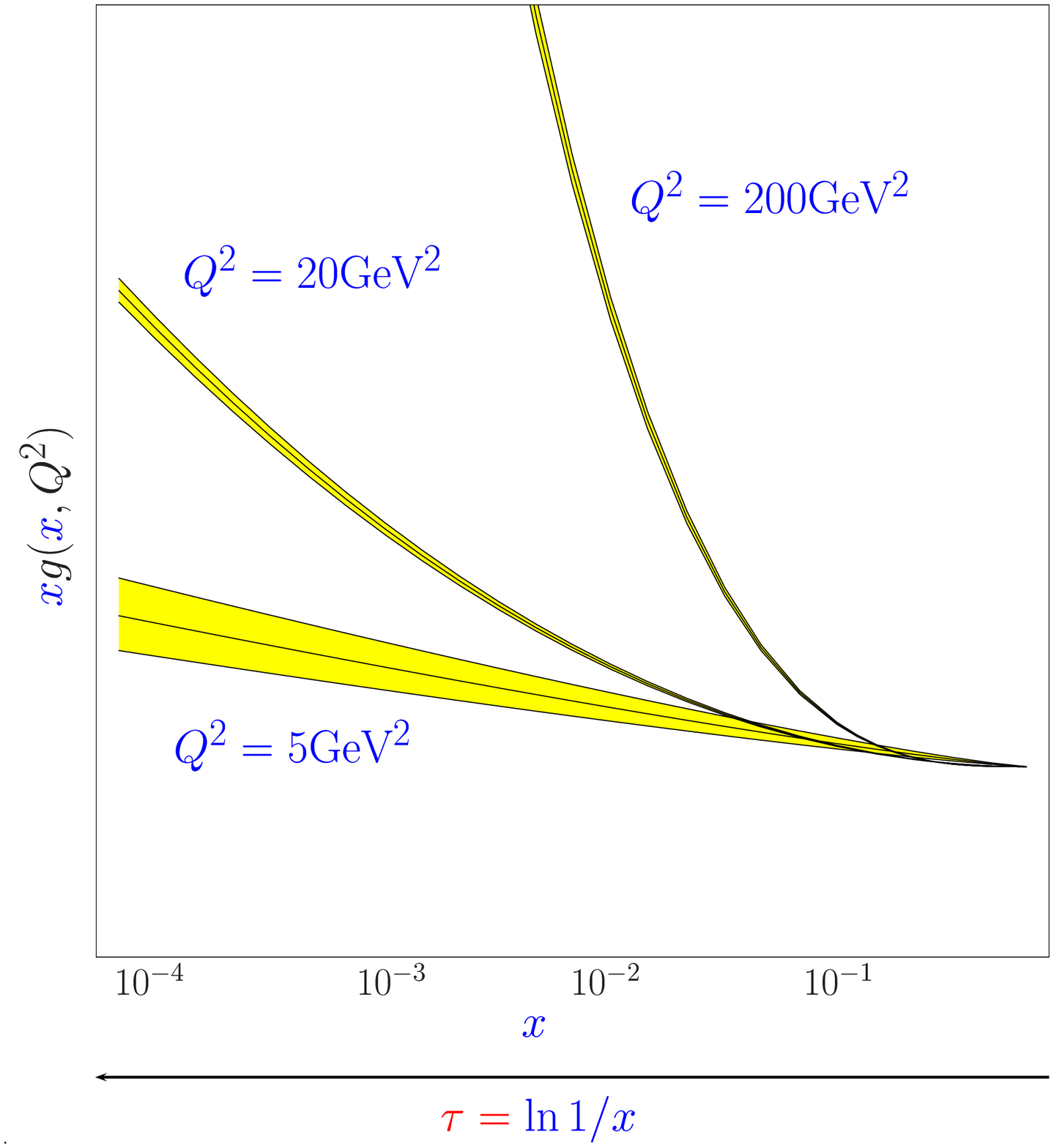}\hspace{2cm}
   \includegraphics[width=6cm]{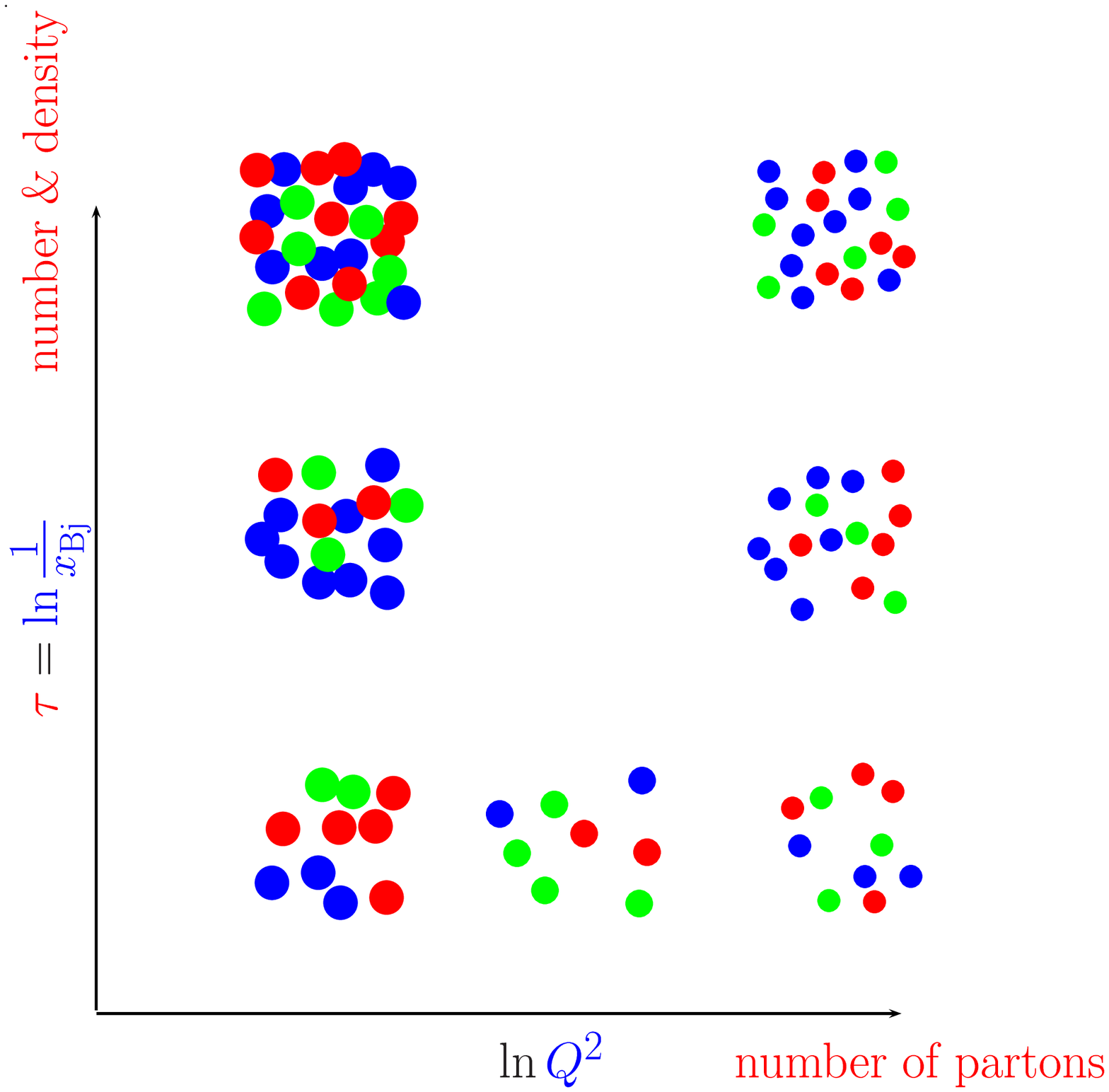}
  \caption{\small Gluon distributions grow towards small $x$. To the left: 
    qualitative growth as from typical DGLAP fits of DIS at HERA ($e
    A$) for different $Q^2$.  To the right: Growth of gluon (parton)
    numbers with $Q^2$ and $\ln(1/x)$. Going to sufficiently small $x$
    at fixed $Q^2$ partons start to overlap. }
  \label{fig:x-Q-plane-density}
\end{figure}

Thus we are considering a collision of a virtual photon that imparts a
(large) space-like momentum $q$ ($Q^2:=-q^2$ is large) on a nucleus of
momentum $p$. At large energies the only other relevant invariant is
Bjorken $x$ with $x:= \frac{Q^2}{2 p.q}$. A leading twist analysis of
DIS employing an OPE based on the largeness of $Q^2$ allows us to
express the the cross section entirely in terms of two point
functions, the quark and gluon distributions. At the same time it
becomes possible to calculate their $Q^2$ dependence to leading
logarithmic accuracy (summing corrections of the form $(\alpha_s \ln
Q^2)^n$) by solving the DGLAP equations. This amounts to summing the
well-known QCD ladder diagrams.  This procedure is self consistent and
perturbatively under control as long as we stay within the large $Q^2$
region. Starting from a large enough $Q^2$ and going to larger values,
we find that the objects we count with the quark and gluon
distributions increase in number but they stay dilute: their ``sizes''
follow the transverse resolution scale $1/Q$. At small $x$, however,
the growth of the gluon distribution is particularly pronounced and
any attempt to resum corrections of the form $(\alpha_s \ln(1/x))^n$)
to track the $x$-dependence of the cross sections immediately is faced
with extreme growth of gluon distributions: moving towards small $x$
at fixed $Q^2$ increases the number of gluons of fixed ``size'' $1/Q$,
so that the objects resolved will necessarily start to overlap. This
is depicted in Fig.~\ref{fig:x-Q-plane-density}.
At this point we have long since left the region in which an OPE
treatment at leading twist is meaningful. Instead of single particle
properties like distribution functions we need to take into account
all higher twist effects that are related to the high density
situation in which the gluons in our target overlap. At small $x$ we
are thus faced with a situation in which we clearly need to go beyond
the standard tools of perturbative QCD usually based on a twist
expansion.

There are different ways to isolate the leading contributions to the
cross section at small $x$ in a background of many or dense gluons and
the history of the field is long~\cite{Mueller:1986wy, Mueller:1994rr,
  Mueller:1994jq, McLerran:1994ni, McLerran:1994ka, McLerran:1994vd,
  Ayala:1995kg, Kovner:1995ja, Ayala:1996hx, Kovchegov:1996ty,
  Balitsky:1996ub, Kovchegov:1997pc, Jalilian-Marian:1997xn,
  Jalilian-Marian:1997jx, Jalilian-Marian:1997gr,
  Jalilian-Marian:1997dw, Balitsky:1997mk, Jalilian-Marian:1998cb,
  Mueller:1999wm,Kovchegov:1999ua, Kovner:2000pt, Weigert:2000gi,
  Iancu:2000hn,Ferreiro:2001qy}.
We will use a physically quite intuitive picture first used in the
formulation of the McLerran-Venugopalan model, a formulation that was
originally given in the infinite momentum frame of the nuclear target.
Probing this target at small $x$ implies a large rapidity difference
$\tau=\ln(1/x)$ between projectile and target, corresponding to a
large relative boost factor $e^\tau$.  Knowing from the above
considerations that the interaction will be dominated by gluonic
configurations we try to isolate the boost enhanced ones. Looking at
the gluon field strength of the target for that purpose we find that
$F^{+ i}$ components are boost enhanced, while all others are left
small and can, for our purposes, be set to zero. At the same time we
find a strong Lorentz contraction, which we take to be in $x^-$
direction, and a time dilation, correspondingly in $x^+$ direction.
With such a field strength tensor there we are left with only one
important degree of freedom which, by choice of gauge, can be taken to
be the + component of a gauge field. Taking into account time dilation
and Lorentz contraction the gauge field can then be written as
\begin{equation}
\label{eq:aplus}
\begin{split}
  A =& b+\delta A
\qquad \mbox{with} \qquad
  b^{i,-}=  0,\quad b^+=\beta(\boldsymbol{x})\delta(x^-)\ .  
\end{split}
\end{equation}
where the leading contribution $b^+$ is $x^+$ independent and Lorentz
contracted to a $\delta$-function. Both of these are to be taken to be
true with a resolution imposed by the rapidity separation
$\tau=\ln(1/x)$.  Note that mathematically we can always trade a gauge
field that has only a single component for a path ordered exponential
along the direction that picks up this component:
\begin{equation}
  \label{eq:Udef}
  b^+=i(\partial^+U)U^\dagger \hspace{1cm} U={\sf P}
  \exp -i \int\!\!dz^- b^+(z^-,{\bm{x}},0)
\end{equation}
If multiple eikonal interactions are relevant as the high energy
nature of the process would suggest, we expect these path ordered
exponentials to be the natural degrees of freedom.

\begin{figure}[htbp]
  \centering
  \subfigure[]{
       \includegraphics[height=2.6cm]{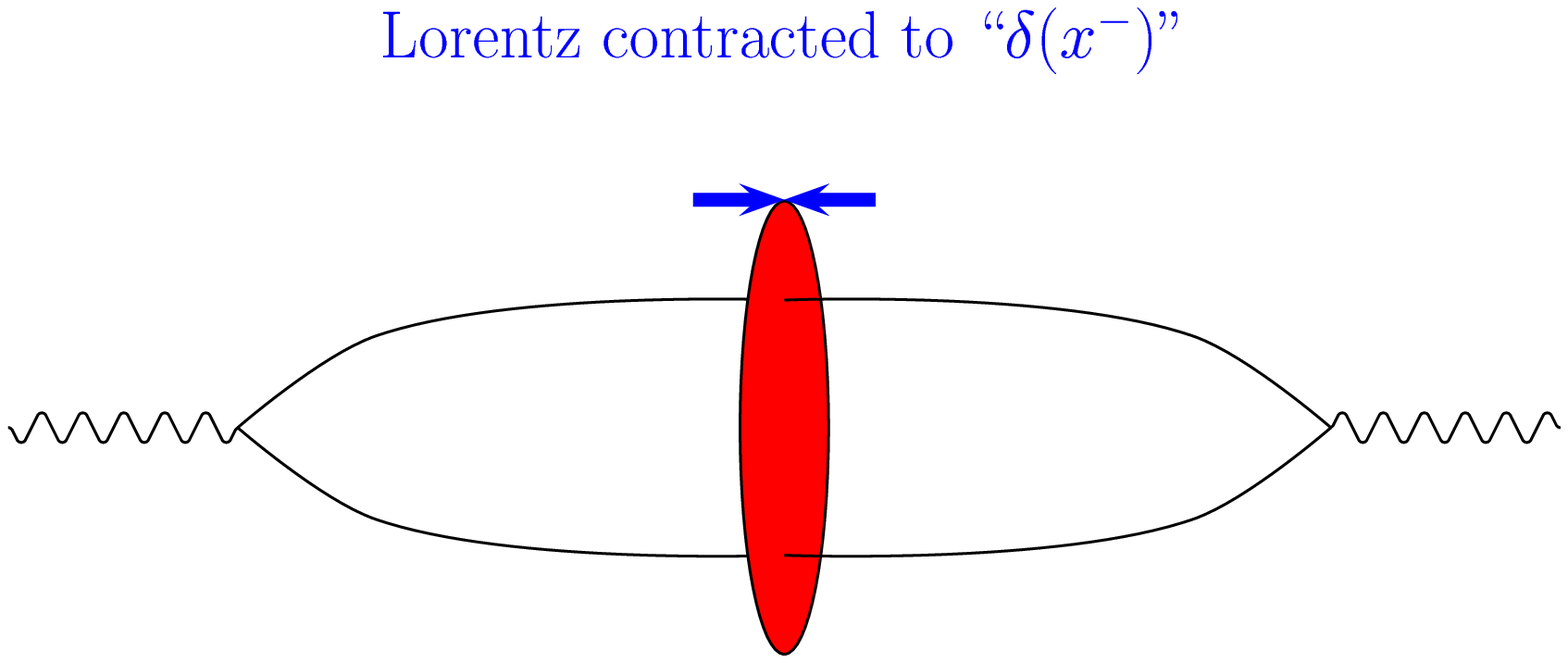} 
       \label{fig:small-x-dis-geom-inf}}
\hfill
  \subfigure[]{
      \includegraphics[height=2.6cm]{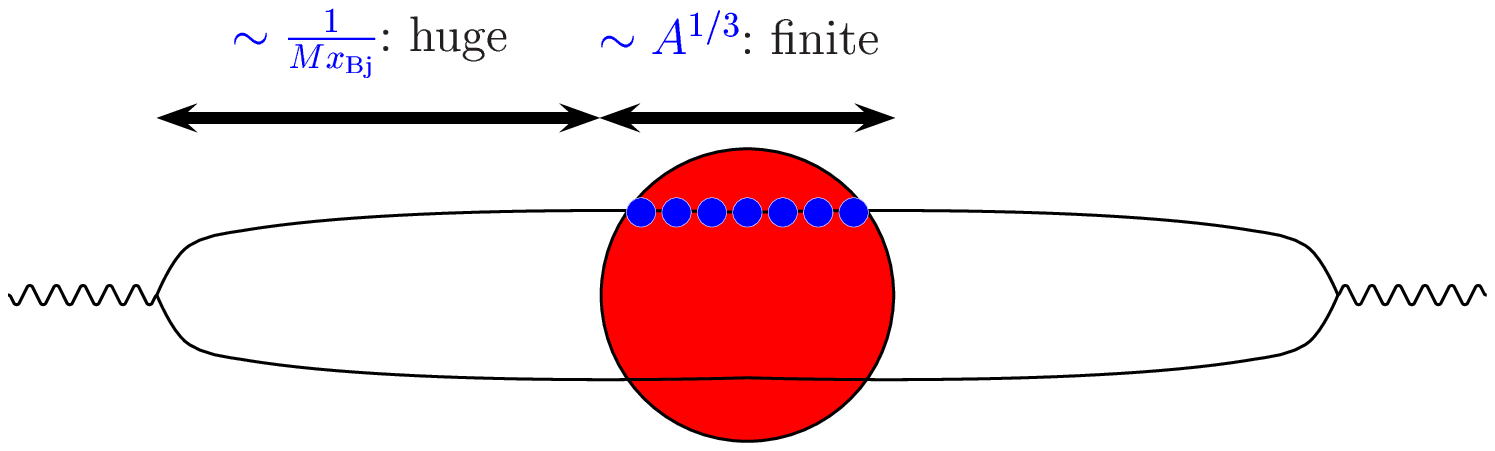}
      \label{fig:small-x-dis-geom-rest}}
  \caption{\small $q\Bar q$ pairs interacting with a nuclear target 
    at small $x$. (a) shows the situation in the targets infinite
    momentum frame with fully Lorentz-contracted target fields, (b)
    shows the situation in the target rest frame in which the
    $\gamma^* q\Bar q$ vertex is far outside the target at a distance
    proportional to $1/x = e^\tau$. }
  \label{fig:small-x-dis-geom}
\end{figure}

Ignoring the small fluctuations $\delta A$ for a moment we have a
picture in which the $\gamma^*$ cross section arises from a diagram in
which the photon splits into a $q\Bar q$ pair which then interacts
with the background field. Due to the $\delta$ like support of $b^+$
the $q\Bar q$ are not deflected in the transverse direction during
that interaction. This just reflects the largeness of the longitudinal
momentum component of these partons at small $x$.  This is shown in
Fig.~\ref{fig:small-x-dis-geom-inf}. To be sure, the physics content
is not frame dependent although it is encoded differently in, say the
rest frame of the target. There, we see neither Lorentz contraction
nor time dilation. However the scale relations are preserved: the
photon splits into a $q\Bar q$ pair far outside the target and its
$p^-$ is so large that typical $x^+$ variations of the target are
negligible during the interaction. As a consequence the probe is not
deflected in the transverse direction, picking up any multiple
interactions with (gluonic) scattering centers as it punches though
the target. This is shown in Fig.~\ref{fig:small-x-dis-geom-rest}.

That multiple interactions are of relevance immediately becomes
obvious once we try to calculate such diagrams with a background field
method. To this end we calculate the diagram shown in
Fig.~\ref{fig:small-x-dis-geom} in the background of a field of the
type shown in Eq.~\eqref{eq:aplus} ignoring the small fluctuations.
This immediately leads to the expression
\begin{equation}
  \label{eq:dipole-cross}
  \sigma_{\mathrm{DIS}}(x_{\mathrm{Bj}},Q^2) = \text{\cal Im}\ 
  \begin{minipage}[c]{3cm}
    \includegraphics[height=.8cm]{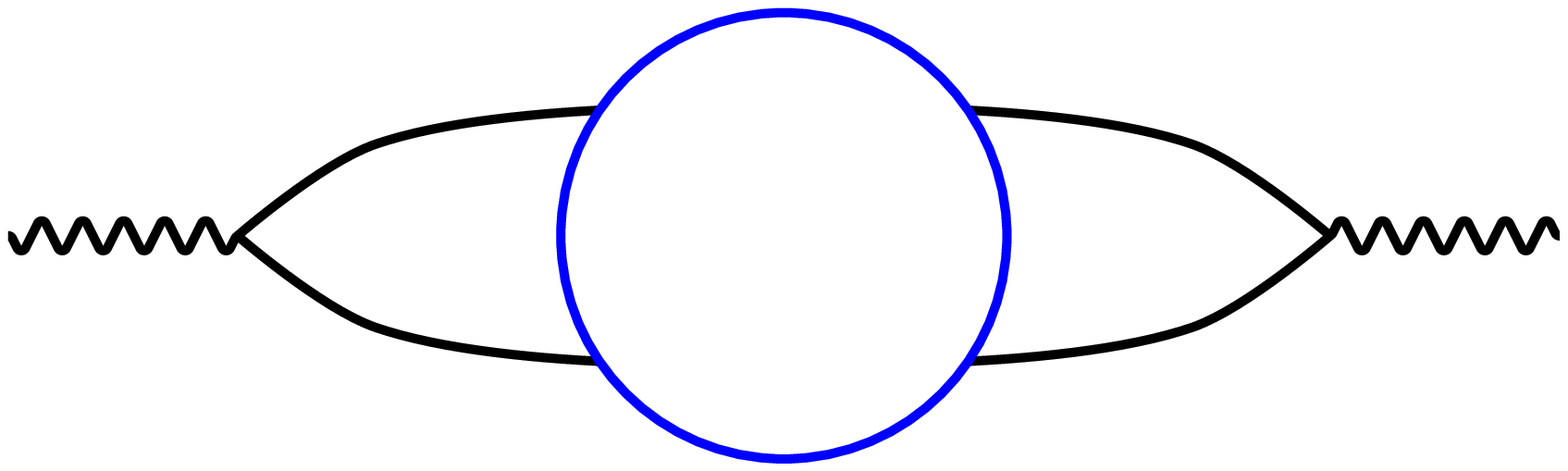}
  \end{minipage}
  =\int\!\!d^2 r 
  \, \vert \psi^2\vert(r^2 Q^2) \hspace{.1cm}
  \int d^2b \ \langle\frac{\tr(1-U_{\bm{x}} U^\dagger_{\bm{y}})}{N_c}\rangle
\end{equation}
where $\bm{r}=\bm{x}-\bm{y}$ corresponds to the transverse size of the
$q\Bar q$ dipole and $\bm{b}=(\bm{x}+\bm{y})/2$ to its impact
parameter relative to the target.  As shown, the leading small $x$
contribution factorizes into a wave function part (the factor $\vert
\psi^2\vert(r^2 Q^2)$) and the dipole cross section part
\begin{equation}
  \label{eq:sigma-dipole}
  \sigma_{\text{dipole}}(\tau=\ln(1/x),r^2) := \int d^2b \ 
  \langle\frac{\tr(1-U_{\bm{x}} U^\dagger_{\bm{y}})}{N_c}
  \rangle_{\tau=\ln(1/x)}
\end{equation}
The former contains the $\gamma^* q\Bar q$ vertices and can,
alternatively to the background field method, be calculated in QED.
They turn out to be linear combinations of Bessel functions $K_{0,1}$
corresponding to the polarizations of the virtual photon.  It contains
all of the direct $Q^2$-dependence of the cross section.

As anticipated above, the dipole cross section part carries all
the interaction with the background gluon fields in terms of the path
ordered exponentials. The averaging $\langle \ldots \rangle_x$ is
understood as an average over the dominant configurations at a
given $x$. It contains all the information about the QCD action and
the target wave functions that are relevant at small $x$.

It is fairly clear from the above discussion, that the isolation of
the leading $b^+$ contribution that defines this average is resolution
dependent idea: as we lower $x$ additional modes, up to now contained
in $\delta A$ of Eq.~\eqref{eq:aplus} will take on the features of
$b^+$. They will Lorentz contract and their $x^+$-dependence will
freeze. Accordingly the averaging procedure will have to change. If
we write the average as
\begin{equation}
  \label{eq:average}
  \langle \ldots \rangle_\tau = 
  \int D[b^+] \ldots W_\tau[b^+]
\hspace{.5cm}\text{or, equivalently}\hspace{.5cm}
  \langle \ldots \rangle_\tau = 
  \int \Hat D[U] \ldots \Hat Z_\tau[U]\ ,
\end{equation}
the weights $ W_\tau[b^+]$ or $\Hat Z_\tau[U]$ will, by necessity be
$\tau=\ln(1/x)$ dependent.

Indeed, while the weights themselves do contain nonperturbative
information, the change of these weights with $\tau$ can be calculated
to leading log accuracy, i.e.\ to accuracy $\alpha_s \ln(1/x)$.
Assuming we know, say $\Hat Z_\tau[U]$ at a given $\tau_0$, this is
done in a Wilson renormalization group manner by integrating over the
fluctuations $\delta A$ around $b^+$ between the old and new cutoffs
$\tau_0$ and $\tau_1$. Taking the limit $\delta\tau=\tau_1-\tau_0\to
0$ we then get a renormalization group equation for $\Hat Z_\tau[U]$.
 
\begin{figure}[htbp]
  \centering
  \includegraphics[height=12cm]{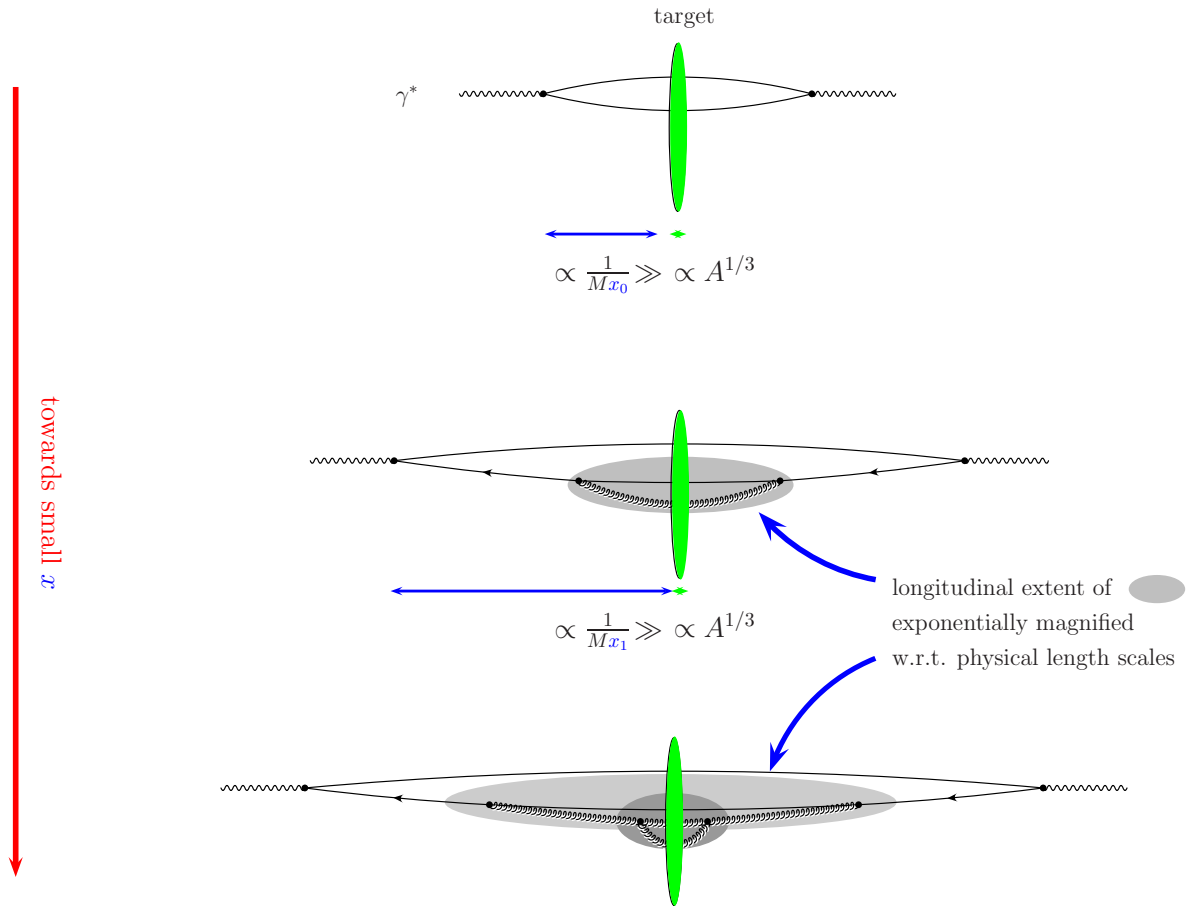}
  \caption{\small RG corrections to the average over background field 
    configurations. The shaded areas contain the contributions
    subsumed into the averaging procedure at successive values of
    $\tau=\ln(1/x)$.}
  \label{fig:RG-corrections}
\end{figure}

Because we integrate over $\delta A$ in the background of arbitrary
$b$ of the form Eq.~\eqref{eq:aplus}, the equation treats the
background field exactly to all orders and captures the all nonlinear
effects also in its interaction with the target.  The resulting RG
equation is functional equation for $\Hat Z_\tau[U]$ that is nonlinear
in $U$. It sums corrections the the leading diagram shown in
Eq.~\eqref{eq:dipole-cross} in which additional gluons are radiated
off the initial $q\Bar q$ pair as shown schematically in
Fig.~\ref{fig:RG-corrections}. All multiple eikonal scatterings inside
the target (the shaded areas) are accounted for.

The result of this procedure is the
Jalilian-Marian+Iancu+McLerran+Weigert+Leonidov+Kovner equation~\cite{
  Jalilian-Marian:1997xn, Jalilian-Marian:1997jx,
  Jalilian-Marian:1997gr, Jalilian-Marian:1997dw,
  Jalilian-Marian:1998cb, Kovner:2000pt, Weigert:2000gi,
  Iancu:2000hn,Ferreiro:2001qy} [the order of names was chosen by Al
Mueller to give rise to the acronym JIMWLK, pronounced ``gym walk''].
This equation and its limiting cases will be discussed in detail in
the next section.

Let us close here with a few phenomenological expectations for the key
ingredient of our phenomenological discussion, the dipole cross
section. Clearly, we expect the underlying expectation value of the
dipole operator, 
\begin{equation}
  \label{eq:N-def}
N_{\tau\ \bm{x y}}:=\langle\frac{\tr(1-U_{\bm{x}}
  U^\dagger_{\bm{y}})}{N_c} \rangle_{\tau=\ln(1/x)}  
\end{equation}
to vanish outside the target, so that the impact parameter integral
essentially samples the transverse target size and thus scales roughly
with $A^{2/3}$.  For central collisions, i.e.\ deep inside the target
we expect the dependence on the dipole size $\bm{r}=\bm{x}-\bm{y}$ to
interpolate between zero at small $\bm{r}$ and one at large $\bm{r}$.
This is the idea of color transparency.  The simplest model to fit
this requirement (which should be good where edge effects can be
expected to be small, i.e.\ for large nuclei) would be to parametrize
the $\bm{b}$-dependence with a $\theta$-function and have the $\bm{r}$
dependence interpolate between $0$ and $1$ in a way that the
transition is characterized by a single scale $Q_s$ usually called the
saturation scale.

In such a model, the $\bm{b}$ integral will only provide a
normalization factor. The idea that the scale in the $\bm{r}$
dependence carries at the same time all the $\tau=\ln(1/x)$-dependence
according to
\label{eq:GB+W-scaling-ansatz}
\begin{equation}
  \label{eq:BK-scaling-0}
  N_{\tau\ \bm{x y}} = N( (\bm{x}-\bm{y})^2 Q_s(\tau)^2 )
\end{equation}
has actually been used by Golec-Biernat and
W{\"u}sthoff~\cite{Golec-Biernat:1998js, Golec-Biernat:1999qd,
  Stasto:2000er} to provide a new scaling fit of HERA $F_2$ data that
is remarkably good. This is particularly true in view of the fact that
we are precisely not talking about large nuclear targets. A slight
generalization of this scaling ansatz has also been used to connect
DIS data on nuclei and protons based on the simple idea underlying the
McLerran-Venugopalan model that $Q_s$ should increase like
\begin{equation}
  \label{eq:Qs-A}
  (Q_s^2)_A = (Q_s^2)_p\ A^{1/3}
\end{equation}
as the additional gluons encountered in additional nucleons on a
straight line path through a larger target add in additional color
charges that decorrelate eikonal scatterers on shorter and shorter
distances. Despite the fact that the data do not really cover $x$- and
$Q^2$-ranges where such scaling arguments should reliably hold, the
agreement is surprisingly good.\footnote{This fit was attempted by the
  authors of the present paper together with Andreas Freund and
  Andreas Sch{\"a}fer with the idea of checking how far presently
  available data failed to scale -- the results were much closer to
  scaling than expected} It would appear that the data available show
a remarkable degree of scaling even in cases that are at best at the
edge of regions where we would expect the underlying ideas to fully
apply.

In the next section will discuss evolution equation and their relation
to scaling features as discussed above. As the ideas there were relying
on central collisions we will find that the evolution equations lead
to such features there. Indeed they turn out not to be valid in very
peripheral regions.

\section{A short review of the evolution equations}
\label{sec:short-revi-evol}

After the qualitative introduction in the previous section, we will
simply state the JIMWLK equation and then discuss its key features and
limiting cases in order to connect its properties with the
phenomenological discussion above. The JIMWLK equation reads
\begin{equation}
  \label{eq:JIMWLK}
  \partial_\tau \Hat Z_\tau[U]=
  -\frac{1}{2} i\nabla^a_{\bm{x}} \chi^{a b}_{\bm{x y}} i\nabla^a_{\bm{y}}  
  Z_\tau[U] 
\end{equation}
and was first cast in this compact form in~\cite{Weigert:2000gi}.
Again, $U_{\bm{x}}$ are the Wilson line variables describing the
kinematically enhanced degrees of freedom, $\nabla^a_{\bm{x}}$ are
functional version of the left invariant vector fields (c.f. Lie
derivatives) acting on $U_{\bm{y}}$ according to
$i\nabla^a_{\bm{x}}U_{\bm{y}}=U_{\bm{x}} t^2\delta^{(2)}_{\bm{x y}}$
(for more details see~\cite{Weigert:2000gi}) and
\begin{equation}
  \label{eq:chidef}
   \chi^{a b}_{\bm{x y}}:= \frac{\alpha_s}{\pi^2}
   \int\!\! d^2\!z \ K_{\bm{x z y}} 
   \Big[(1-\Tilde U^\dagger_{\bm{x}} \Tilde U_{\bm{z}})(1-\Tilde U^\dagger_{\bm{z}}\Tilde U_{\bm{y}})\Big]^{a b} \hspace{1cm}  
   K_{\bm{x z y}} = \frac{(\bm{x}-\bm{z}).(\bm{z}-\bm{y})}{%
     (\bm{x}-\bm{z})^2(\bm{z}-\bm{y}^2)} \ .
\end{equation}
[The ``$\sim$'' indicates the adjoint representation.]  $\Hat{Z}_\tau[U]$
is the ``weight functional'' that determines correlators $O[U]$ of $U$
fields according to
\begin{equation}
  \label{eq:corrs}
  \langle O[U] \rangle_\tau := \int\! \Hat{D}[U]  O[U] Z_\tau[U]
\end{equation}
$\Hat{D}[U]$ in Eq.~(\ref{eq:corrs}) is a functional Haar-measure.

Eq.~(\ref{eq:JIMWLK}) itself is of the form of a (generalized)
functional Fokker-Planck equation. In particular the operator on its
right hand side,
\begin{equation}  \label{eq:HFP}
  H_{\text{\tiny FP}} :=\frac{1}{2} i\nabla^a_{\bm{x}}\, \chi^{a b}_{\bm{x y}}\, i\nabla^a_{\bm{y}}
\end{equation}
is indeed self adjoint and positive definite and may therefore be
called the Fokker-Planck Hamiltonian of the small $x$ evolution.

As a functional equation, Eq.~(\ref{eq:JIMWLK}) is equivalent to an
infinite set of equations for $n$-point correlators of $U$ and $U^\dagger$
fields in any representation of $SU(N_c)$. These will form a coupled
hierarchy due to the $U$-dependent nature of $H_{\text{\tiny FP}}$.
The evolution equation for a given correlator of $U$-fields is easily
obtained by multiplying both sides of Eq.~(\ref{eq:JIMWLK}) and
integrating the result with a functional Haar measure. Using the
self-adjoint nature of $H_{\text{\tiny FP}}$ and the fact that the Lie
derivatives do not interfere with the Haar measure we immediately
obtain
\begin{equation}
  \label{eq:op-JIMWLK}
  \partial_\tau \langle O[U] \rangle_\tau =  \langle (-\frac{1}{2}i\nabla^a_{\bm{x}} \chi^{a b}_{\bm{x y}} i\nabla^a_{\bm{y}}  O[U]) \rangle_\tau \ .
\end{equation}
The only step remaining in the derivation of the evolution equation
for $O[U]$ is to evaluate the Lie derivatives as they act on $
O[U]$. The nonlinear nature of $H_{\text{\tiny FP}}$ implies that
the expectation value on the r.h.s. will contain more $U$ fields than
the l.h.s. and thus a new type of correlator of $U$ fields. To
determine $\langle O[U] \rangle_\tau$ we thus will need also the evolution
equation of this new operator in which the same mechanism will couple
in still further $U$ fields. Continuing the process we end up with an
infinite hierarchy of equations.

As an example for $ O[U]$ with immediate phenomenological
relevance we consider the two point function of the dipole operator
\begin{equation}
  \label{eq:dipole}
 N_{\tau,\bm{x}\, \bm{y}}:=  \langle \Hat N_{\bm{x y}} \rangle_\tau \hspace{2cm} 
 \Hat N_{\bm{x}\, \bm{y}} := \frac{\tr(1 - U_{\bm{x}}^\dagger U_{\bm{y}})}{N_c}
\end{equation}
[Below we will often suppress the explicit $\tau$ dependence.]
Eq.~(\ref{eq:op-JIMWLK}) immediately leads to
\begin{equation}
  \label{eq:pre-BK}
  \partial_\tau \big\langle \Hat N_{\bm{x y}} \big\rangle_\tau  = \frac{\alpha_s N_c}{2\pi^2}\int\!\! d^2 z \frac{(\bm{x}-\bm{y})^2}{%
    (\bm{x}-\bm{z})^2(\bm{z}-\bm{y})^2}\big\langle   
  \Hat  N_{\bm{x x}}+ \Hat  N_{\bm{z y}}- \Hat N_{\bm{x y}}- \Hat  N_{\bm{x z}}\, \Hat  N_{\bm{z y}}\big\rangle_\tau 
\end{equation}
Clearly the r.h.s. depends on a 3-point function containing a total of
up to 4 $U^{(\dagger)}$ factors, that in general does not factorize into a
product of two 2-point correlators (or linear combinations thereof):
\begin{equation}
  \label{eq:4-point-nonfac}
 \big\langle  \Hat  N_{\bm{x z}}\, \Hat  N_{\bm{z y}}\big\rangle_\tau  
\neq \big\langle  \Hat  N_{\bm{x z}} \big\rangle_\tau \big\langle \Hat   N_{\bm{z y}}\big\rangle_\tau \ .
\end{equation}
To completely specify the evolution of $\langle \Hat N_{\bm{x y}}
\rangle_\tau$ we therefore also need to know $\big\langle \Hat
N_{\bm{x z}}\, \Hat N_{\bm{z y}}\big\rangle_\tau $. The latter, in its
evolution equation, will couple to yet higher n-point functions and
thus we are faced with one of the infinite hierarchies of evolution
equations that are completely encoded in the single functional
equation~(\ref{eq:JIMWLK}).

Because of the fact that Eq.~(\ref{eq:JIMWLK}) encodes information on
the evolution of {\em all} possible n-point functions any generic
statement that can be derived without reference to a particular
correlator is of particular importance.  Interestingly enough, and
despite its functional nature it allows us to deduce already some of
the main features of the evolution process. First of all,
Eq.~(\ref{eq:JIMWLK}) is of Hamiltonian form with a
semi-positive-definite Hamiltonian and describes a diffusion process
in which $\tau$ plays the r{\^o}le of (imaginary) time. From this observation
alone, we get very strong statements without any additional work: the
eigenmodes of this Hamiltonian, $Z_{n,\tau}[U]$, possesses positive real
eigenvalues $\lambda_n$ and a generic solution will evolve as a
superposition of such modes in the form
\begin{equation}
  \label{eq:eigenmodes}
  Z_{\tau}[U]=\sum\limits_n a_n e^{-\lambda_n(\tau-\tau_0)} Z_{n,\tau_0}[U]
\end{equation}
where the sum is a formal sum over all modes in the spectrum.  If
there is an eigenvalue $\lambda_0=0$, this will be the only one not
damped away exponentially with $e^{-\lambda_n\tau}$ at large $\tau$.
Indeed $Z_{\tau,\lambda=0}[U]=1$ furnishes such a candidate and it can
be argued to be unique~\cite{Weigert:2000gi}. This solution features
as a fixed point of the evolution equation and, since all other
contributions naturally die away, it is attractive. This remains true
although $H_{\text{FP}}$ is scale invariant and the spectrum, labelled
by $n$ in the above, is continuous.

At this fixed point, all correlators will be fully determined by what
remains in the averaging procedure after $Z_{\tau}[U]\to 1$, namely
the Haar measure.  This however will average everything to zero unless
the transverse locations of the $U_{\bm{x}_i}$ factors in the
correlator coincide at least pairwise and for {\em each} distinct
transverse point remaining there is a singlet piece in the
decomposition of the $U$ factors into irreducible representations.
This immediately implies that all correlation lengths at $\tau=\infty$
vanish: evolution must erase them at $\tau$ grows. What does that mean
for our standard example the two point or dipole correlator?  As we
have discussed, we expect it to interpolate between zero at small
separation and one at large separation, with the transition say at a
scale $1/Q_s(x_0)$ at some finite $x_0$.  $1/Q_s(x_0)$ is the
correlation length visible in this correlator. As we lower $x$ the
correlation length will shrink to zero and we expect the following
qualitative trend in its $x$ evolution:
\begin{figure}[htbp]
  \centering
  \includegraphics[width=7cm]{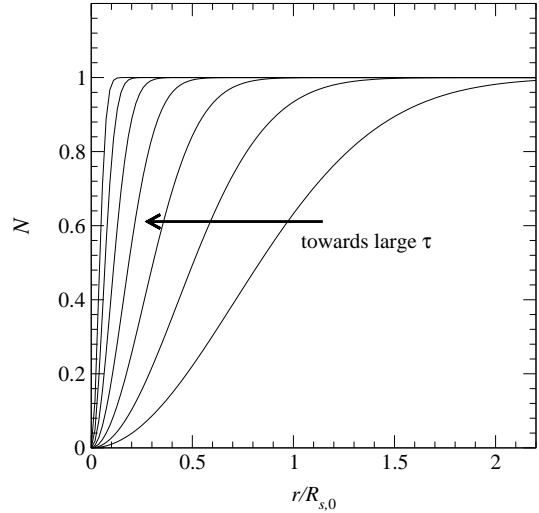}
  \caption{\small Generic evolution trend for a one scale dipole 
    correlator $N_{\bm{x y}}$ plotted against $\vert
    \bm{x}-\bm{y}\vert$ as $x\to 0$: the curves move towards the left
    as $\tau=\ln(1/x)$ increases, the asymptotic form being a step
    function at the origin }
  \label{fig:generic-evol}
\end{figure}

There is, in principle, a way to implement Eq.~(\ref{eq:JIMWLK})
numerically. Eq.~(\ref{eq:JIMWLK}) is a Fokker-Plank equation and as
such can be rewritten as a Langevin equation~\cite{Weigert:2000gi,
  Blaizot:2002xy}. The latter can be treated numerically with a
lattice discretization of transverse coordinate space. We will discuss
this at length in Sec.~\ref{sec:jimwkl-bk-equations}.

It is useful to first study the qualitative properties of this
equation in a simplified setting, the only approximation scheme known
that preserves the main qualitative features depicted in
Figure~\ref{fig:generic-evol}. This turns out to be a $1/N_c$
expansion.  Viewed from this perspective Eq.~\eqref{eq:4-point-nonfac}
turns into
\begin{equation}
  \label{eq:4-point-fac-Nc}
 \big\langle  \Hat  N_{\bm{x z}}\, \Hat  N_{\bm{z y}}\big\rangle_\tau  
= \big\langle  \Hat  N_{\bm{x z}} \big\rangle_\tau \big\langle \Hat   N_{\bm{z y}}\big\rangle_\tau +{\cal O}(\frac{1}{N_c})\ .
\end{equation}
This is in fact general: dropping the corrections naturally factorizes all
$n$-point functions into $2$-point functions and higher correlators
decouple for example from Eq.~\eqref{eq:pre-BK}. The remainder turns
into
\begin{equation}
  \label{eq:BK}
  \partial_\tau N_{\bm{x y}} = \frac{\alpha_s N_c}{2\pi^2}\int\!\! d^2 z \frac{(\bm{x}-\bm{y})^2}{%
    (\bm{x}-\bm{z})^2(\bm{z}-\bm{y})^2}\Big(  
  N_{\bm{x x}}+ N_{\bm{z y}}-N_{\bm{x y}}- N_{\bm{x z}}\, N_{\bm{z y}}\Big)\ ,
\end{equation}
\label{page:BK-intro}
where the $\tau$ dependence of the dipole function $N$ is implicitly
understood.  Eq.~\eqref{eq:BK} was independently derived by Ian
Balitsky from his ``shock wave formalism''~\cite{Balitsky:1996ub,
  Balitsky:1997mk} that is completely summarized by
Eq.~(\ref{eq:JIMWLK}) and by Yuri Kovchegov from Mueller's dipole
formalism for the case of large nuclei~\cite{Kovchegov:1999ua}.
Indeed the whole infinite coupled hierarchy of equations represented
by Eq.~(\ref{eq:JIMWLK}) is reduced in content to this one equation for one 
single function $N_{\bm{z y}}$, called the Balitsky-Kovchegov- or
BK-equation.\footnote{Equations for higher correlators of course still
  exist. They appear as sums with the BK-equation repeated in all
  terms in a way that follows the factorization of the left hand side.
  For example $\partial_\tau \big\langle \Hat N_{\bm{x z}}\, \Hat
  N_{\bm{z y}}\big\rangle_\tau = (\partial_\tau N_{\bm{x z}})N_{\bm{z
      y}} + N_{\bm{x z}}\, (\partial_\tau N_{\bm{z y}}) = (\text{right
    hand side of BK})_{\bm{x z}} N_{\bm{z y}} + N_{\bm{x
      z}}(\text{right hand side of BK})_{\bm{z y}}$, up to terms of
  order $1/N_c$. }

The main feature of this equation is that it remains consistent with
an interpretation of $1-N_{\bm{x y}}$ as a correlator of Wilson lines
$\langle \tr( U^\dagger_{\bm{x}} U_{\bm{y}}) \rangle /N_c$ in that the
nonlinearity keeps $N$ between zero and one. Loosely speaking, this is
due to a cancellation of the linear and nonlinear terms in $N$ within
Eq.~(\ref{eq:BK}). In the low density limit, where the $N^2$ term is
negligible we are left with the BFKL equation in a space time setting
as first demonstrated in~\cite{Balitsky:1997mk}.  The solution for the
equation {\em without} the nonlinear term is indeed of the BFKL form
\begin{equation}
  \label{eq:N-BFKL-sol}
  N_{\tau, \bm{x y}} = \int \frac{d\nu}{2\pi^2} d^2r_0\
  ((\bm{x}-\bm{y})^2)^{-\frac{1}{2}+i\nu} 
  e^{(\tau-\tau_0)\chi(\nu) 
  } 
 (\bm{r}_0^2)^{-\frac{1}{2}-i\nu} 
  \ N_{\tau_0,\bm{r}_0}
\end{equation}
with arbitrary initial condition $N_{\tau_0,\bm{r}_0}$ and $\chi(\nu)$
the BFKL eigenvalue function. This solution is known to grow and
spread equally to smaller and larger sizes. The first property limits
its range of applicability: as soon as $N$ has grown enough as to
``switch on'' the nonlinearities, the result Eq.~\eqref{eq:N-BFKL-sol}
will no longer hold. The second shows that the linear equation spreads
into the infrared. Where that happens we have do abandon it
altogether. If the nonlinearity is active at short enough distances,
i.e. within the perturbative domain we expect two effects: the growth
of $N$ should be tamed and the infrared should never affect the
evolution.

Numerical studies~\cite{Golec-Biernat:2001if, Braun:2000wr} using a
momentum space representation of the BK equation with a boundary
condition localized at a single momentum have shown that this is
indeed the case. Moreover the solutions have been found to approach a
scaling regime in which the coordinate and $x$ dependence are coupled
according to
\begin{subequations}
\label{eq:BK-scaling-ansatz}
\begin{equation}
  \label{eq:BK-scaling}
  N_{\tau,\bm{x y}} = N( (\bm{x}-\bm{y})^2 Q_s(x)^2 )\ .
\end{equation}
There the saturation scale $Q_s(x)$ grows like a power of $1/x$.
However, the initial conditions and hence the dipole correlators in
these simulations do not satisfy the ``color transparency'' boundary
conditions referred to above.

That this simply means that scaling behavior is according to
Eq.~\eqref{eq:BK-scaling} is quite generic was shown
in~\cite{Iancu:2002tr}. There the authors show that the existence of
scaling solutions of the type in Eq.~\eqref{eq:BK-scaling}, now with
with physics inspired spatial ``color transparency'' boundary
conditions of the type
\begin{equation}
  \label{eq:BK-boundary}
  N(0) = 0 \hspace{1cm}\text{and}\hspace{1cm} N(\infty)=1\ ,
\end{equation}
\end{subequations}
also leads to power like growth of the saturations scale according to
\begin{equation}
  \label{eq:scaling-Q_s}
  Q_s(x) = \Big(\frac{x_0}{x} \Big)^\lambda Q_0
\end{equation}
If it is possible to satisfy all of the requirements in
Eqns.~\eqref{eq:BK-scaling-ansatz}, the value of the constant
$\lambda$ is determined self-consistently through the equation together
with the shape of $N$.\footnote{The argument in\cite{Iancu:2002tr}
  does not prove the {\em existence} of such solutions, it merely
  translates scaling ansatz + boundary conditions into a particular
  form of the $x$ dependence of the saturation scale given in
  Eq.~\eqref{eq:scaling-Q_s}. Indeed the argument would go though even
  without the nonlinearity. In this case, however, the equation is
  closely related to the BFKL equation, with BFKL eigenvalues and
  power like eigenmodes. The generic solution then is given in
  Eq.~\eqref{eq:N-BFKL-sol} and does not possess solutions with the
  properties in Eq.~\eqref{eq:BK-scaling-ansatz}.} That the equation
indeed possesses such solutions has, up to now only been demonstrated
numerically and form part of our results below. From our simulations
it would appear that that single scale initial conditions that satisfy
Eq.~\eqref{eq:BK-boundary} typically evolve into scaling solutions of
the form Eq.~\eqref{eq:BK-scaling}.

In principle it is possible to extend the scaling argument
of~\cite{Iancu:2002tr} to solutions of the JIMWLK equation although
there one is talking about an infinite hierarchy of different
$n$-point functions and the claim that this would naturally fit into a
one scale description would sound even more arbitrary than for the BK
case. The reason why such solutions are important is that they appear
to represent the small $x$ asymptotic behavior for a large class of
initial conditions. Unlike the qualitative features imposed by the
infinite energy fixed point of Eq.~(\ref{eq:JIMWLK}), this statement
is much more directly amenable to experimental verification: if the
initial conditions realized in nature put us not too far from this
asymptotic behavior, we find a unique and universal feature in the
$x$-dependence of any possible target for which the BK equation yields
a reasonable description of the zero impact parameter contributions.

It remains important to understand the conditions under which such
behavior can be expected:

First note, that the scaling behavior Eq.~\eqref{eq:BK-scaling-ansatz}
with growing saturation scale $Q_s$ is just a special case of the
generic behavior shown in Fig.~\ref{fig:generic-evol}.  Whether we are
in a purely one scale situation or in a setting with more than one
scale enters the scale that controls the transition of $N_{\tau,\bm{x
    y}} \to 1$ at large $\bm{r}=\bm{x}-\bm{y}$ sets the infrared scale
in the problem as $N$ will not grow beyond $1$. For this to happen we
need large densities of gluons. Where this is guaranteed at some
initial $\tau_0=\ln(1/x_0)$, general fixed point considerations and
explicit simulations show that the IR scales will only get larger as
we continue to evolve towards smaller $x$. Where the initial IR scale
is not perturbative the derivation of the evolution equations is not
valid, however. This limits our description to central impact
parameters off large enough nuclei. 

Besides these facts there remains a lot to be learned about solutions
to the nonlinear evolution equations presented in this section. This
is the goal of this paper.

Obviously, the main advantage of Eq.~(\ref{eq:BK}) over the Langevin
representation of Eq.~(\ref{eq:JIMWLK}) is its simplicity. Although,
even here, no analytical solutions are available, the numerical task
at hand is much more tractable. If we can establish that this equation
allows us to identify most of the qualitative features of the full
evolution, we can use it to explore the main features of small
$x$-evolution an only then resort to a numerical treatment of
Eq.~(\ref{eq:JIMWLK}) in the most interesting regions to get
quantitatively precise results. It is precisely this r{\^o}le in which we
are going to utilize the BK-equation.

Our work consists of two main parts:
\begin{itemize}
\item First, we perform a study of the JIMWLK equation, using its
  Langevin form implemented using a lattice discretization of
  transverse space. 
  
  This is the first coordinate space treatment of either the JIMWLK or
  BK equations and we are the first to show {\em directly} that the
  solutions approach the fixed point as qualitatively predicted
  in~\cite{Weigert:2000gi}, c.f.\ Fig.~\ref{fig:generic-evol}.
  
  At the same time we will be able to confirm IR safety of the JIMWLK
  equation via numerical simulations.
  
  Still at finite $\tau$ we will see that scaling solutions emerge
  that are characterized by a saturation scale $Q_s(x)$, a feature
  already found earlier in the context of the
  BK-equation~\cite{Braun:2000wr,Golec-Biernat:2001if}.
  
  The initial conditions chosen in this study are almost free of any
  $N_c$-violations.  We measure how these fare under full JIMWLK
  evolution and find that the equations appear to neither enhance nor
  dampen the extent of these.
  
  To complement our study of IR stability we also carefully study the
  UV or continuum limit. We will show that diffusion into the UV is
  not tamed. Recalling the properties of the BFKL equation this does
  not come as a complete surprise. However, we will show that for some
  quantities like the evolution rate
  $\lambda_\tau:=\partial_\tau\ln(Q_s(\tau))$ this fact makes a
  reliable continuum extrapolation based on JIMWLK a task of enormous
  computational cost.
  
  Due to the nearly perfect $N_c$-factorization properties associated
  with the initial conditions used we are then able to complement our
  JIMWLK studies with BK simulations implemented with same type of
  lattice UV regulator. This allows for a more thorough study of the
  continuum limit and reveals even closer agreement of JIMWLK and BK
  behavior for the classes of initial condition chosen. We will show
  that the active phase space in both cases extends to up to 7 orders
  of magnitude above $Q_s$, the physical scale of the problem. In such
  situations one is typically forced to resum large logarithmic
  corrections to get quantitatively reliable results. This then leads
  us to the second part:
  
\item Motivated by results found in connection with the BFKL equation
  we will identify left out running coupling effects as the main
  culprits for the largeness of UV phase space.  We perform
  simulations with a coupling running with the size of the parent
  dipole as $\alpha_s(1/(\bm{x}-\bm{y})^2)$ and establish that this
  removes the extraneous phase space and thus resums large phase space
  corrections. Only with this resummation in place the main
  contributions to evolution originate at distances at and around
  $1/Q_s(x)$ and the coupling indeed is of the order of
  $\alpha_s(Q_s(x)^2)$ as anticipated in the underlying physics
  picture.
  
  $\Lambda_{\text{QCD}}$ and scale breaking enter our simulation as a
  consequence of this step. This, at least conceptually, breaks the
  asymptotic, geometric scaling limit of the equations. Numerical
  evidence nevertheless shows that the solutions still converge to an
  approximate asymptotic form that shows almost perfect geometric
  scaling, in agreement with~\cite{Braun:2003un}. Nevertheless we will
  show that a clear modification of the $x$-dependence of $Q_s$
  emerges, away from the simple fixed coupling $(x_0/x)^\lambda$-form.
  $\lambda_{\text{eff}}$ (suitably defined, see below) picks up
  pronounced $\tau$ dependence and the numerical values change
  drastically: we find typically a factor of two to three compared to
  the fixed coupling situation. While the necessity of the inclusion
  of running coupling effects had been seen already in estimates of
  evolution rates and numerical
  simulations~\cite{Golec-Biernat:2001if, Braun:2003un} as well as
  phenomenological fits of HERA data~\cite{Gotsman:2002yy},
  size and conceptual importance of these corrections has not been
  discussed from the viewpoint taken here.

\end{itemize}

\section{JIMWLK and BK equations at fixed coupling}
\label{sec:jimwkl-bk-equations}

\newcommand{\be}{\begin{equation}}
\newcommand{\ee}{\end{equation}}
\newcommand{\bx}{{\bm{x}}}
\newcommand{\by}{{\bm{y}}}
\newcommand{\bz}{{\bm{z}}}
\newcommand{\br}{{\bm{r}}}
\newcommand{\fr}{\frac}

\subsection{Simulating the JIMWLK equation}
\label{sec:numer-results-from}

To be able to solve the JIMWLK equation directly we need a numerical
method to implement its content. Such a method offers itself after
translating the Fokker-Planck equation to a Langevin equation. Such an
approach has already been suggested in~\cite{Weigert:2000gi} and a
derivation has been given in~\cite{Blaizot:2002xy} where we refer the
reader for details. For the interested, we provide a short exposition
of the key steps of a derivation that highlights the close connection
between Langevin equations and path-integrals in
App.~\ref{sec:from-fokker-planck}.

To prepare translation into Langevin form we
follow~\cite{Weigert:2000gi} and write the RG again in standard form
\begin{align}
  \label{RG-new-standard}
  \frac{\partial}{\partial\tau} \Hat Z[U] 
& =
 -
 \nabla^a_{\bm{x}}\Big[\frac{1}{2}\nabla^b_{\bm{y}} \Hat\chi^{a b}_{\bm{x y}} 
-\big(\frac{1}{2}\nabla^b_{\bm{y}} 
\Hat\chi^{a b}_{\bm{x y}} \big)\Big]\Hat Z[U]
\end{align}
and define 
\begin{equation}
  \label{eq:newsigdef}
  \Hat\sigma^a_{\bm{x}}:= \frac{1}{2}\nabla^b_{\bm{y}} 
\Hat\chi^{a b}_{\bm{x y}}= - i \big(\frac{1}{2}\frac{\alpha_s}{\pi^2}\int\!\!d^2z 
\frac{1}{({\bm{x}}-{\bm{z}})^2} \Tilde \tr( \Tilde t^a \Tilde U_{\bm{x}}^\dagger  
\Tilde U_{\bm{z}})\Big)
\end{equation}
(the ``$\sim$'' indicating adjoint matrices and traces). 

We then abandon the description in terms of weight functionals $\Hat
Z_\tau$ in favor of one in terms of ensembles of fields which are
governed by the corresponding Langevin equations.

Explicitly, to calculate any observable
$O[U]$ of the fields $U$ we write
\begin{equation}
  \label{eq:ensemble-average}
  \langle O[U] \rangle_\tau = \int\!\Hat D[U] O[U] \Hat Z_\tau[U] 
  \approx 
  \frac{1}{N} \sum\limits_{U\in {\sf E}[\Hat Z_\tau]}  O[U]
\end{equation}
where, separately at each $\tau$, the sum is over an ensemble ${\sf
  E}[\Hat Z_\tau]$ of $N$ configurations $U$ whose members were
created randomly according to the distribution $\Hat Z_\tau$. Clearly,
for $N\rightarrow\infty$, the ensemble and $\Hat Z_\tau$ contain the
same information.

The Langevin equation then schematically\footnote{This is a continuous
  time version of the equation that strictly speaking is not unique.
  The path-integral derivation shown in
  App.~\ref{sec:from-fokker-planck} makes it clear that we are to take
  a ``retarded'' prescription here in which the derivative on the
  l.h.s\ is taken as a finite difference and the fields on the r.h.s.\ 
  are determined at the previous time step.} reads
\begin{equation}
  \label{eq:Langevin}
  \partial_\tau\, [U_{\bm{x}}]_{i j} 
  = [U_{\bm{x}} i t^a]_{i j} \Big[\int\!\! d^2y\, 
  [{\cal E}_{\bm{x} \bm{y}}^{a b}]_k 
  [\xi^b_{\bm{y}}]_k+\Hat\sigma^a_{\bm{x}}\Big]
\end{equation}
where
\begin{equation}
  {\cal E}^{ab}_{\bx\by} = \left(\fr{\alpha_s}{\pi^2}\right)^{1/2} 
  \fr{(\bx-\by)_k}{(\bx-\by)^2} 
    [ 1 - \Tilde U^\dagger_\bx \Tilde U_\by ]^{ab} 
\end{equation}
is the ``square root'' of $\chi$, 
$\chi^{ab}_{\bx\by} = {\cal E}^{ac}_{\bx\bz} {\cal E}^{cb}_{\bz\by}$,
and $\xi$ are independent Gaussian random variables with
correlators determined according to
\begin{equation}
  \label{eq:etacorr}
  \langle\ldots \rangle_\xi = \int\!D[\xi]\, (\ldots)\,
  e^{-\frac{1}{2}\xi \xi}
\ .
\end{equation}
Note the factor of $i$ which is essential to render this an
equation for an infinitesimal change of an element of $SU(N_c)$. In
fact the components of $\omega^a_{\bm{x}}:=\Big[\int\!\!
d^2y\, [{\cal E}_{\bm{x} \bm{y}}^{a b}]_k
[\xi^b_{\bm{y}}]_k+\Hat\sigma^a_{\bm{x}}\Big]$
can be directly interpreted as the ``angles'' parametrizing a local
gauge transformation in transverse space.

It is of particular importance to note that the possibility to
formulate the stochastic term via a completely decorrelated Gaussian
noise $\xi$, that is to say with $\langle \xi^{a,i}_{\bm{x}}
 \xi^{b,j}_{\bm{y}} \rangle = \delta^{a b} \delta^{i j} 
\delta^{(2)}_{\bm{x}\bm{y}}$, reduces the numerical
cost for a simulation like this considerably.

In our practical implementation, we discretize the Langevin equation
in Eq.~\eqref{eq:Langevin} using a regular square lattice of $N^2$
sites (volume $(Na)^2$, where $a$ is the lattice spacing), with
periodic boundary conditions. In dimensionless units ($a = 1$) and
defining a rescaled, discrete, evolution time
\begin{equation}
  \label{eq:resc-evol-tau}
  s := \frac{\alpha_s}{\pi^2}\tau
\end{equation}
the Langevin equation becomes 
\begin{equation} 
U(\bx; s + \delta s) =
U(\bx;s) \exp[i t^a \omega^a(\bx;s)] 
\end{equation}
where
\begin{eqnarray}
  \omega^a(\bx,s) &=& \sqrt{\delta s} 
        \sum_{\by} K_i(\bx-\by) 
        [ 1 - \tilde U^\dagger(\bx;s)\tilde U(\by;s)]^{ab} \xi_i^b(\by) 
        \nonumber\\
  & - &  \delta s \sum_{\by} S(\bx-\by) \fr12 \Tilde \tr 
        [i \Tilde t^a \tilde U^\dagger(\bx,s) \tilde U(\by,s)]. 
  \label{eq:omega}  
\end{eqnarray}
Here 
\begin{equation} 
   K_i(\br) = \fr{\br_i}{\br^2} \,,
   \hspace{2cm}
   S(\br) = \fr1{\br^2},
\end{equation}
 and $\xi$ is a Gaussian noise with 
$
  \langle \xi_i^a(\bx)\xi_j^b(\by) \rangle = 
        \delta_{ij}\delta^{ab}\delta_{\bx,\by}.
$
$K_i(\br)$ and $S(\br)$ are taken to be $ =0$ if $\br =0$.

In practical numerical work we have to carefully consider the
following issues:

{\bf Boundary conditions:}~ The sum over the volume ($\by$) above must
be defined in conjunction with the boundary conditions used.  In
general, the periodic boundary conditions minimize the effects of the
finite volume and makes the system, on average, translationally
invariant.  Because of the long range of the evolution kernel the
distance between points $\bx$ and $\by$ must be evaluated taking the
periodicity into account.  This can be done in several non-equivalent
ways; we choose one which evaluates the distance $(\bx - \by)$ to be
the shortest one while taking the periodicity into account.  This is
achieved by modifying the argument of the functions $K_i$ and $S$ as
follows: 
\begin{equation}
  \label{eq:periodic}
  (\bx - \by) \mapsto (\bx - \by) + \bm{n} N,
\end{equation}
where $n_i \in \{-1,0,1\}$ is chosen to minimize the absolute value of the
r.h.s. of Eq.~\eqref{eq:periodic}.  
This procedure cuts the range of the evolution kernel to be $\le N/2$,
i.e. we take the volume integral over a single periodic copy of the
volume.

Let us note here that the boundary condition is significant only if
the saturation radius $R_s \sim 1/Q_s$ is of order system size.  As we
shall see below, the boundary conditions (or the system size) have a
negligible effect on the evolution as long as $1/Q_s \ll N$.

{\bf Fourier acceleration:} ~ Due to the non-locality of the evolution
operator, one time step evolution of the $N\times N$ system using
Eq.~\eqref{eq:omega} requires
$\propto N^4$ numerical operations.  This is 
impractical except for the smallest volumes.
However, we note that Eq.~\eqref{eq:omega} can be decomposed into
convolutions as follows:
\begin{equation}
  \omega^a = \sqrt{\delta s} \left[ C(K_i,\xi_i^a)
       - (\tilde U^\dagger)^{a b} C(K_i,\tilde U^{a b} \xi^b_i) \right]
       - (\delta s) \left[ (t^a)^{b c} (\tilde U^\dagger)^{c d} 
          C(S,\tilde U^{d b})\right].
\end{equation}
The convolution can be evaluated with Fourier transformation:
\begin{equation}
   C(A,B)_{\bx} = \sum_{\by} A(\bx-\by) B(\by) = 
        {\cal F}^{-1}[ {\cal F}(A) {\cal F}(B)].
\end{equation}
Thus, using FFTs the cost can be reduced to
$N^2 \log N$, which makes a huge difference.  Nevertheless, the cost
is still large: there are 192 normal or inverse transformations in
each time step (note that $K_i$ and $S$ need to be transformed just
once).

{\bf Time discretization:}~ The
discrete time step $(\delta s)$ introduces an error $\propto (\delta
s)^2$ for each update step.  We chose $\delta s$ so that
the largest values of $\omega^a(\bx)$,
which characterizes the change in $U(\bx)$ in a single step,
always remains smaller than $0.3$. This resulted to time steps $\delta
s \sim 5\times10^{-5}$, which is overly conservative; even quadrupling
$\delta s$ had no observable effect on the evolution.

{\bf Spatial continuum limit:}~ It is essential to have control
over the finite lattice cutoff effects.
In the continuum the evolution
equations are completely scale invariant; the physical length scale is
given by the initial conditions.  On the lattice we have 2 new scales,
lattice size $L$ and -spacing $a$ (which equals 1 in dimensionless
units above).  These scales must be chosen so that the physical scale,
saturation scale $R_s$, is safely between them; in any case, the
effects of the finite IR and UV cutoffs must be checked.  In short, it
turns out that the requirement $R_s \ll L$ is easy to satisfy; a
factor of 3--4 is sufficient in order to have a negligible effect
(i.e. there are no IR problems).  On the other hand, we find that
$a$ should be several orders of magnitude smaller than 
$R_s$ for the UV effects to vanish, making the continuum limit
very non-trivial.

{\bf Initial conditions:}~ For the initial condition we choose a
distribution of $U$-matrices with a Gaussian correlation function:
\begin{equation}
   \langle U^\dagger_\bx U_\by\rangle \propto  
        \exp\left(\fr{-(\bx-\by)^2}{4R^2}\right) .
\end{equation}
Scale $R$ here is close to the initial saturation scale $R_s$
(depending on the precise definition of $R_s$).

For a linear (scalar or vector) field it is straightforward to
generate distributions with (almost) arbitrary initial 2-point
correlation functions, $\langle A_0A_\bx\rangle = C_\bx$, using
Fourier transformations: just let $A_{\bm{k}} \propto
\sqrt{C_{\bm{k}}} \xi_{\bm{k}}$, where $\xi_{\bm{k}}$ is white
Gaussian noise, $P(\xi) \propto \exp(-\xi^2)$, and $C_{\bm{k}}$ is
Fourier transform of $C_{\bx}$.  Field $A_\bx$ is obtained with an
inverse Fourier transformation.

In our case the degrees of freedom are elements of SU(3) group and
this method cannot be applied directly.  However, we can utilize the
approximate linearity of small changes in group elements and build up
the desired distributions in short steps.  The method works as
follows: let the desired correlation function be $\langle
U^\dagger_\bx U_0 \rangle = C_\bx$.  We start from initial state
$U_\bx = 1$ and repeat the operation
\begin{equation}
   U_\bx \leftarrow e^{i A_\bx /\sqrt{n}} U_\bx  
\end{equation}
$n$ times, generating new suitably distributed adjoint matrices
$A_\bx=A_\bx^a t^a$ for each step.  
Here $A^a_\bx$ is a Fourier transform of $A^a_{\bm{k}} \propto
(C^A_{\bm{k}})^{1/2} \xi^a_{\bm{k}}$, where $\xi^a_{\bm{k}}$ are independent
Gaussian random variables and $C^A_{\bm{k}}$ is the Fourier transform
of $-\ln C_\bx$ (The use of the logarithm is a trick to give the
correct asymptotics to the correlation function.).

Thus, the matrices $U_\bx$ ``diffuse'' towards the desired state by
random walking along the group manifold.  The number of the steps,
$n$, has to be large enough in order to give us properly randomized
$U$-matrices.  Naturally, the method is only approximate, but the
deviations from the desired correlation function remain small.

We also used another, completely unrelated, method for generating the
initial state: this time we start from a {\em completely random} configuration
of $U_\bx$-matrices $\in$ SU(3), and make the substitution
\begin{equation}
  U_\bx \leftarrow \exp(\fr12 R^2\nabla^2) U_\bx 
  \approx (1 + \fr1{2n} R^2 \nabla^2)^n U_\bx\,,
   \label{eq:smooth}
\end{equation}
for $n$ large enough.  $\nabla^2$ is evaluated as
lattice finite differences.  Thus, the last expression tells us to
``smear'' the matrix $U_\bx$ with its neighbors $n$ times.
For linear fields this procedure
yields exactly Gaussian 2-point function $\langle U_\bx U_by
\rangle \propto \exp[-(\bx-\by)^2/2R^2]$.  This is straightforward to
verify by making a Fourier transformation of Eq.~\eqref{eq:smooth}.
However, because we have to ensure that $U_\bx$ remains on the group
manifold, we have to project $U$ back to SU(3) after each
step.  This is done by finding $U' \in \mbox{SU(3)}$ which
maximizes $\tr( U^\dagger U')$, and substituting $U\leftarrow U'$.
The projection step makes the initial state generation method again
approximate.

Both of these two methods yield initial ensembles which have 
almost Gaussian initial 2-point functions.  Due to performance
reasons we used mostly the first method (the latter method requires
$n \gg L$ in order to really produce long-distance correlations).

\subsection{Numerical results from JIMWLK simulations}

In our analysis we use volumes $30^2$--$512^2$, with 8--20 independent
trajectories (with independent initial conditions) for each volume.
This amount of statistics proved to be sufficient for our purposes;
actually trajectory-by-trajectory fluctuations in the dipole operators
are quite small (i.e. already one trajectory gives a good estimate of
the final result).

\begin{figure}[htbp]
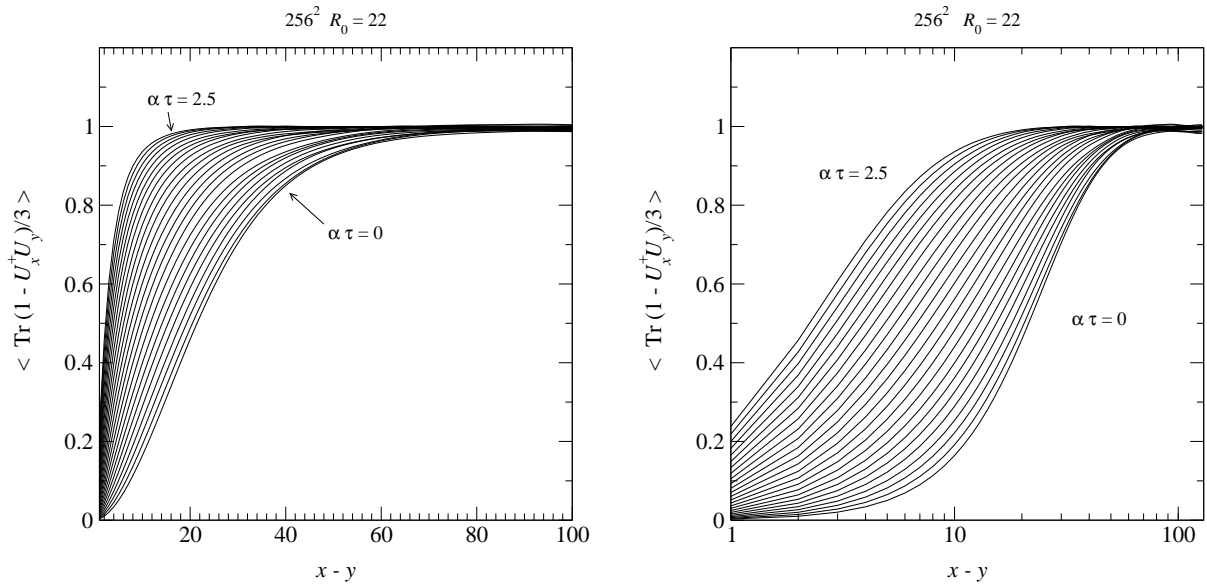

  \centering
  \includegraphics[height=7.7cm]{UUcorr_su3_256b}\hspace{.5cm}
  \includegraphics[height=7.7cm]{UUcorr_su3_256b_log}
  \caption{\small Evolution of the dipole operator from the JIMWLK equation on 
    an $256^2$ lattice, shown on a linear (left) and logarithmic
    (right) distance scale.  The general pattern as predicted from the
    existence of the fixed point and drawn qualitatively in
    Fig.~\ref{fig:generic-evol} shows up clearly. On the right we see
    a scaling form emerge: a stable shape of the curve that merely
    shifts to the left unaltered.}
  \label{fig:jimwlk-evo}
\end{figure}

In Fig.~\ref{fig:jimwlk-evo} we show the evolution of the dipole
operator, averaged over 8 independent trajectories on a $256^2$
lattice, from interval $\alpha_s \tau \equiv \alpha_s \ln(1/x) = 0
\ldots 2.5$.  Qualitatively, we can observe that the dipole operator
soon approaches a specific functional scaling form, where the
shape is preserved but the length scale is shortened under the
evolution.  This is clearly visible on the logarithmic scale plot on
the right: the Gaussian initial condition quickly settles towards the
scaling solution which evolves by moving towards left while
approximately preserving the shape.

From the dipole correlation function we can now measure the evolution
of the saturation length scale $R_s$.  There are several inequivalent
ways to measure $R_s$ from the dipole function; perhaps the most
straightforward and robust method is to measure the distance where the dipole
function reaches some specified value.  This is the approach we adopt
here; precisely, we define $R_s(\tau)$ to be the distance where the
dipole function $N((\bm{x}-\bm{y})^2)=1/N_c \tr\langle 1 -
U^\dagger_\bx U_\by\rangle$ reaches a value $c$:
\begin{equation}
  \label{eq:R_s-def-c}
  N(R_s^2):=c
\end{equation}
where $c$ should not be much smaller than 1.  The precise definition
of course is $c$-dependent -- different choices will lead to a
rescaling of the value of $R_s$ -- and we adopt the convention to set
$c=1/2$.  Such a definition can be used both inside and outside the
scaling region and will encode some physics information in both.

In Fig.~\ref{fig:rs-tau} we show $R_s(\tau)$, measured from various
lattice sizes and initial values for $R_s(\tau = \tau_0)$.  Since the
initial $\tau_0$ is not a physical observable in the sense that we are
not in position to match initial conditions to experiment, we are at
liberty to shift the $\tau$-axis for each of the curves in order to
match the curves together.  In Fig.~\ref{fig:rs-tau} this procedure
clearly gives us a section of the $R(\tau)$ curve, independent of the
system size or initial size.

From $R_s(\tau)$ we obtain the instantaneous evolution speed 
\begin{equation}
\lambda = - \fr{\partial \ln R_s}{\partial\tau}.  
\end{equation}
Note that contrary to $R_s$ this quantity is unique and independent of
the convention for $c$ in Eq.~\eqref{eq:R_s-def-c} {\em inside} the
scaling regime. Outside this uniqueness is lost.

In Fig.~\ref{fig:lambda-r} we show $\lambda$ plotted against $R_s$,
measured from several volumes and initial sizes.  The curves naturally
evolve from right to left, from large $R_s$ towards smaller values.
As opposed to Fig.~\ref{fig:rs-tau}, in this case the curves are fully
determined by the measurements.  Note that for a scale invariant
evolution $\lambda$ should be {\em constant}, independent
of $R_s$.  This is clearly not the case here.
We observe the following features:

\begin{figure}[htbp]
  \centering
  \begin{minipage}[t]{7.8cm}
\includegraphics[height=7cm]{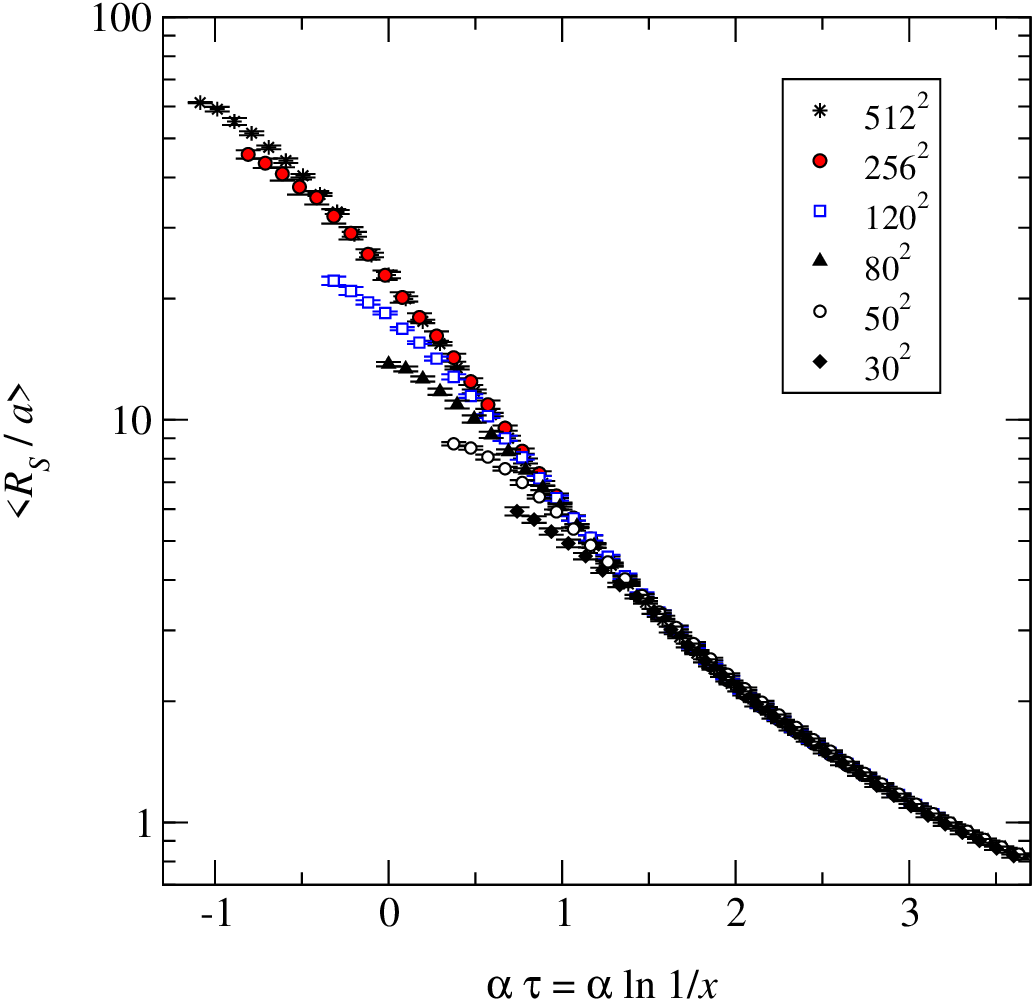}
  \caption{\small The evolution of the saturation scale $R_s$, measured
   from system sizes $30^2$ -- $512^2$.  The curves are shifted
   in $\tau$ to match at large $\tau$.}
  \label{fig:rs-tau}
  \end{minipage}
\hspace{.5cm}
  \begin{minipage}[t]{7.8cm}
\includegraphics[height=7cm]{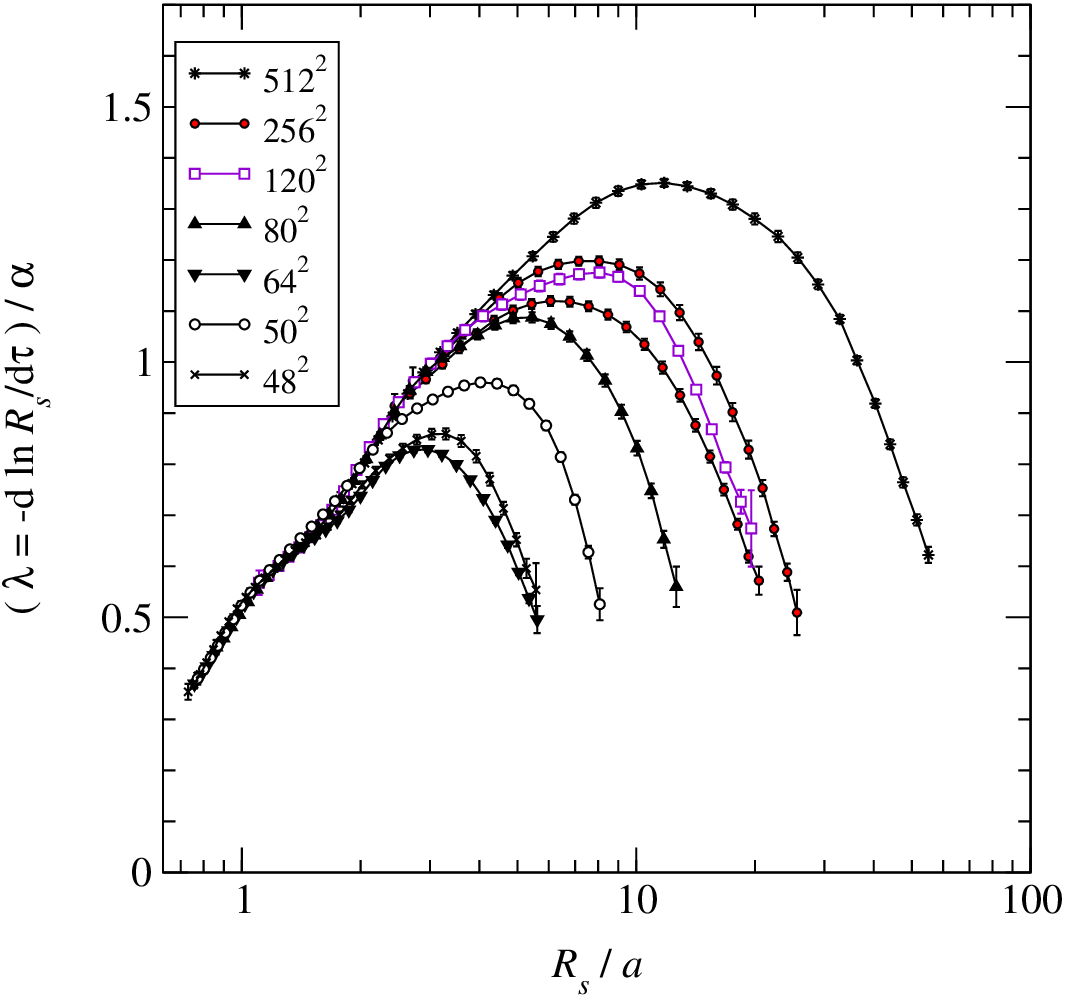}
  \caption{\small The evolution speed $\lambda = - 
    \partial_\tau\ln R_s$ plotted against $R_s$ for various
    system sizes and initial $R_s$.}
  \label{fig:lambda-r}
  \end{minipage}
\end{figure}

\begin{figure}[htbp]
  \centering
  \begin{minipage}[t]{7.8cm}
\includegraphics[width=8cm]{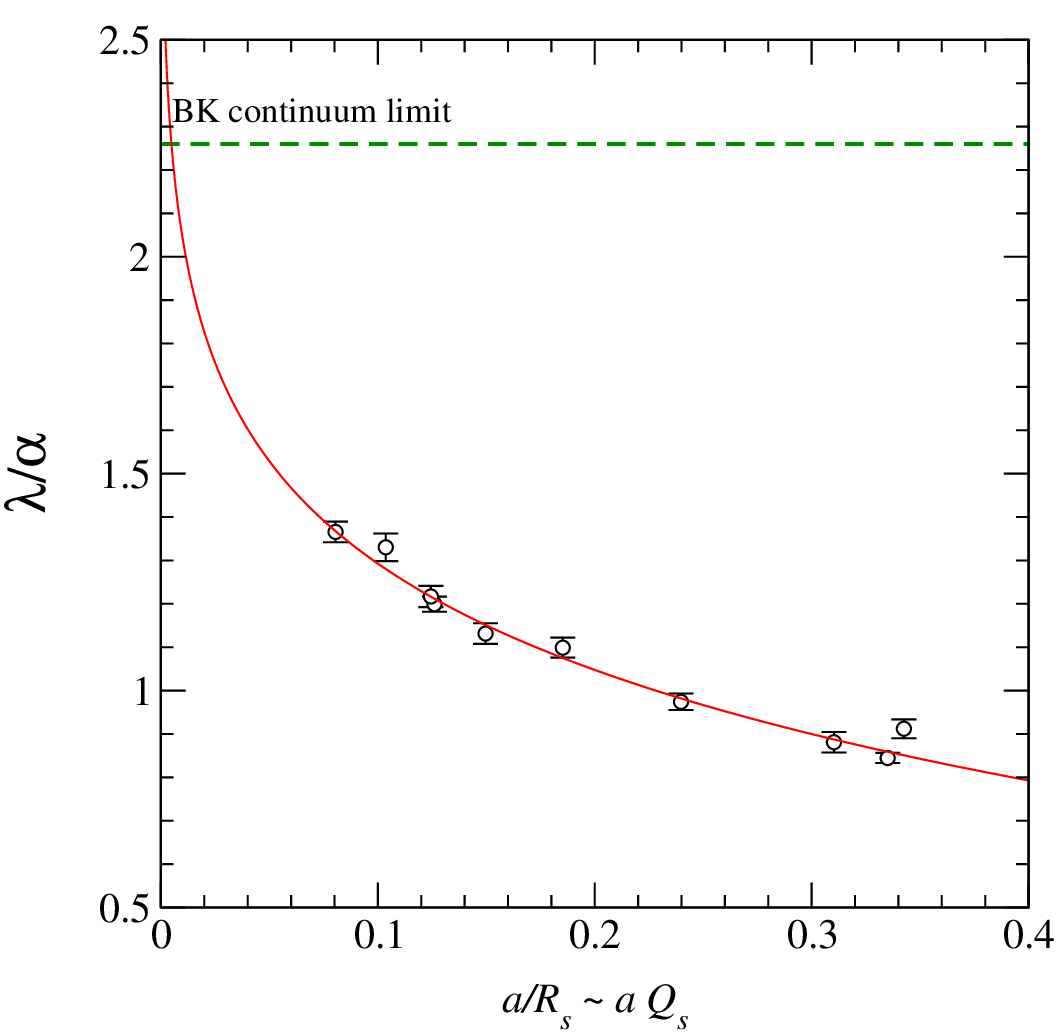}
  \caption{\small Approach to the continuum limit of $\lambda/\alpha$ 
    against $a$  in units of 
    $R_s$. The values are taken from the maxima in
    Fig.~\ref{fig:lambda-r}. The dashed line on the top shows the
    continuum value obtained from a simulation of the BK equation. A
    continuum extrapolation (continuous line) from this alone appears
    unreliable with most of the data away from the steep section.}
  \label{fig:jimwlk-UV-extrap}
  \end{minipage}
\hspace{.5cm}
  \begin{minipage}[t]{7.8cm}
\includegraphics[width=8cm]{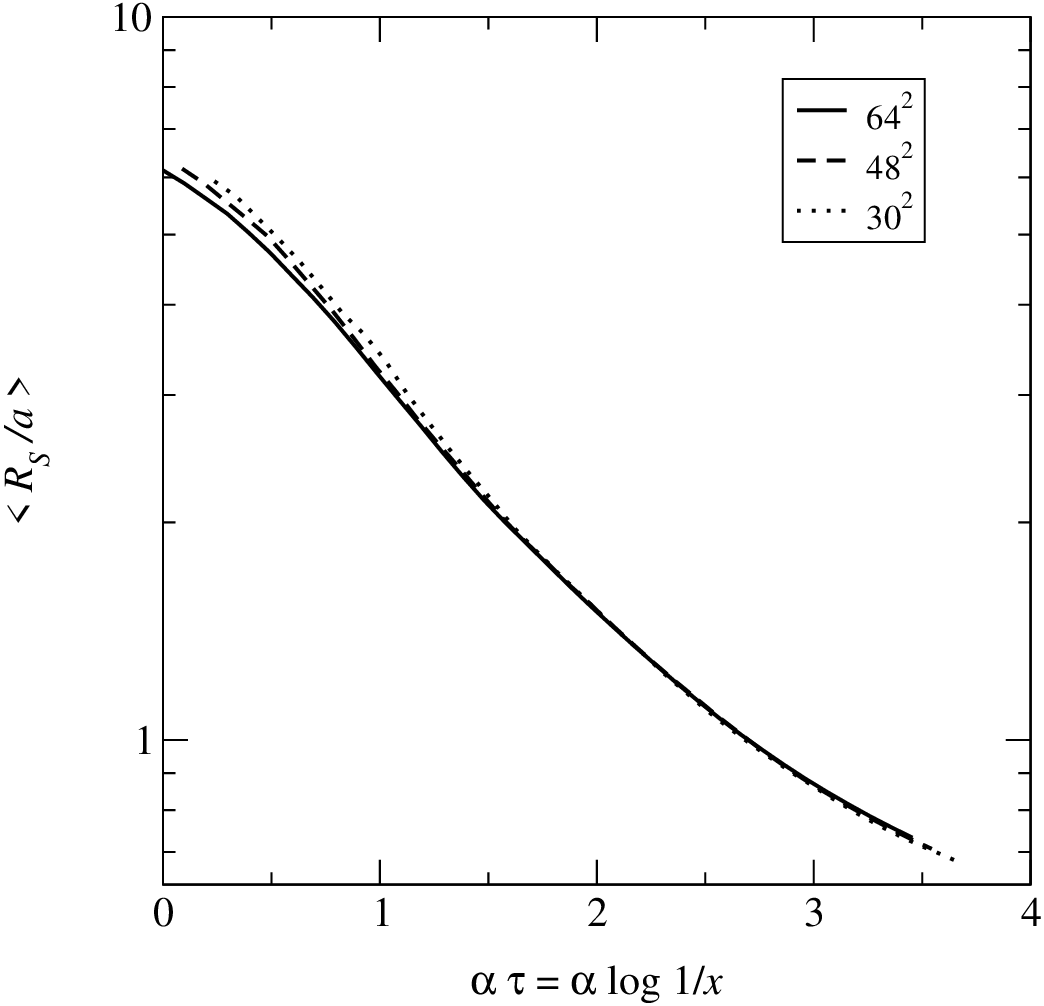}
  \caption{\small The evolution of the saturation scale with different IR
  cutoffs (lattice sizes).  Except for a small initial difference, the
  differences are well within statistical errors.}
  \label{fig:jimwlk-IR}
  \end{minipage}
\end{figure}

\begin{itemize}
\item $\lambda$ starts relatively small but increases rapidly.  We
  interpret this to be due to the settling of the Gaussian initial
  state towards the scaling form.  This change in shape is clearly
  visible in Fig.~\ref{fig:generic-evol}.  Such behavior typical for
  initial state effects.  This effect should be much smaller if the
  initial ensemble would more closely resemble the scaling
  solution.  In these particular simulations, in which we were
  mainly interested in universal features of the evolution equations
  at larger $\tau$, part of the initial changes also arise from the
  fact that the initial $R_s$ may be too large; i.e.  the initial
  state dipole function may be too far away from unity at distances
  $L/2$, the maximum distance on a $L^2$ lattice.

\item At small $R_s$ (large $\tau$), $\lambda(R_s)$ clearly settles on
  an universal curve, which decreases with decreasing $R_s$.  This
  curve is a function of $R_s/a$, i.e. the saturation scale {\em in
    units of the lattice spacing}.  Thus, this is a UV cutoff effect
  which should vanish (curve turn horizontal) when $a\rightarrow 0$
  ($R_s/a \rightarrow \infty$).  However, we are clearly still far
  from this case, even on largest volumes.  In order to obtain a
  continuum limit, one should perform extrapolation $a\rightarrow 0$.
  We do this as follows: first, we note that the maxima of
  $\lambda(R_s/a)$ -curves for each of the runs fall on a universal
  curve when plotted against $a/R_s$ at maximum.  This is shown in
  Fig.~\ref{fig:jimwlk-UV-extrap}.  Linear function does not fit the
  data well; on the other hand, fitting a power law + constant we get
  a reasonable match (shown in the figure).  However, with the
  steepest part of the curve in a region without data it is clear that
  the extrapolation is not safe.  Without a definite functional form
  for the extrapolation our data is not good enough to allow for a
  reliable continuum limit estimate. (Nevertheless, we note that the
  comparison to BK simulation results, discussed below, makes the
  continuum limit in JIMWLK much clearer.)

\item On the other hand, the IR cutoff effects appear to be mild: the
  behavior of $\lambda(R_s)$ curves is largely determined by the
  initial radius $R_s(\tau_0)$, independent of the system size $L$ (so
  far as $L$ is clearly larger than $R_s$).  This is clarified in
  Fig.~\ref{fig:jimwlk-IR}, where evolution of $R_s(\tau)$ is shown
  from volumes $30^2$ -- $64^2$, using the same $R_s(\tau_0) \approx
  6a$.  Only small initial deviations are seen.

\end{itemize}

Thus, the necessity of having a very large hierarchy of the relevant scales
$a \ll R_s < L$ prevents us from making continuum extrapolations.
The cutoff sensitivity problem is also present in the BK equation
(\ref{eq:BK}).  However, the BK equation is much simpler to study
numerically than the full JIMWLK equation: firstly, there is only a
single degree of freedom, a scalar field instead of a SU(3) matrix
field, and secondly the evolution equation is fully deterministic,
instead of statistical.  This enables us to study much larger range of
volumes and initial conditions with reduced errors.

\begin{figure}[htbp]
  \centering
  \includegraphics[width=7.5cm]{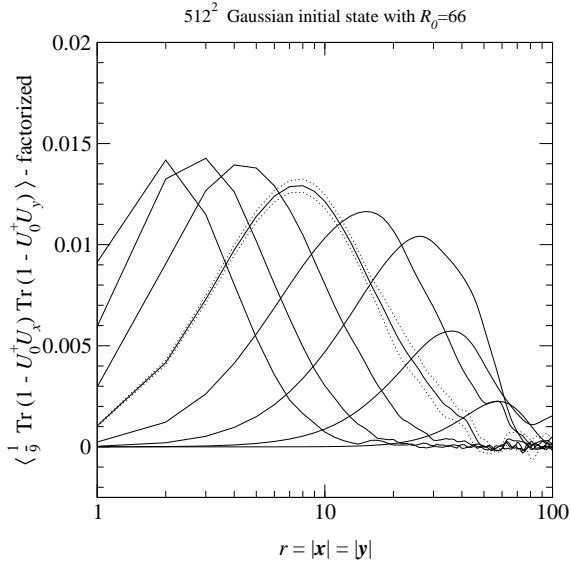}
  \caption{\small Factorization violation for a $4$-field correlator. Typical 
    error size is indicated by dashed lines for a curve in the middle.
    They are generically small and indicate that the large $N_c$
    limit, i.e.  the BK equation will be a good approximation as long
    as the initial condition does not contain strong $1/N_c$
    corrections.  }
  \label{fig:Nc-violations}
\end{figure}

Moreover, under certain conditions, the BK equation can be expected to
be a fairly good approximation of the JIMWLK equation: in
Fig.~\ref{fig:Nc-violations} we show the difference of 4-field
correlators
\begin{equation}
 \langle \tr( U_\bx^\dagger U_\by) 
        \tr(U_\by^\dagger U_\bz) \rangle 
- \langle \tr( U_\bx^\dagger U_\by)\rangle
        \langle \tr(U_\by^\dagger U_\bz) \rangle.
\end{equation}
For concreteness, we chose the points so that $(\bx - \by) \parallel
\bm{e}_1$, $(\by - \bz) \parallel \bm{e}_2$ and $|\bx -\by| = |\by -
\bz|$.  The natural magnitude for these correlators is $\sim 1$.
Thus, given the initial conditions we used, we see that the
correlators cancel to a 1--2 \% accuracy, with the violations staying
roughly at the same size throughout the $\tau$ interval covered.
This seems to indicate that for the type of initial conditions used,
the BK equation should give a good approximation of the salient
features of the evolution. In particular we should be able to refine
our understanding of the UV extrapolation and augment
Fig.~\ref{fig:jimwlk-UV-extrap} with results from BK simulations in
which we also use a lattice representation for transverse space to
parallel our treatment of the JIMWLK equation.

\subsection{BK and JIMWLK compared}
\label{sec:bk-jimwlk-compared}

In order to study the cutoff-dependence and to facilitate the
comparison with JIMWLK simulations we study the BK equation with two
different numerical approaches.

First, we discretize Eq.~(\ref{eq:BK}) on a regular square lattice with 
periodic boundary conditions as in JIMWLK.  Using the translation
invariance, we can substitute $N_{\bx\by} \rightarrow N_{\bx - \by}$,
and set, for example, $\by = 0$ in Eq.~(\ref{eq:BK}), obtaining
\begin{equation}
    \partial_\tau N_\bx = \frac{\alpha_s N_c}{2\pi^2}\int\!\! d^2 z 
        \frac{\bx^2}{(\bx-\bz)^2 \bz^2}\Big(  
  N_{\bx-\bz} + N_\bz - N_\bx - N_{\bx-\bz}\, N_\bz\Big)\ .
\end{equation}
This equation can be readily simulated on a square lattice.  The
evolution kernel again can be written in terms of convolutions, and
can be efficiently evaluated using FFTs.   We again make the choice
that when $\bz = 0$ or $\bz = \bx$, the integral kernel evaluates to
zero.  The periodicity is taken into account analogously to the
JIMWLK equation.

The second method we use has been developed
by~\cite{Golec-Biernat:2001if}, similar results had earlier been
obtained in~\cite{Braun:2000wr}.  By first Fourier transforming the
evolution equation (\ref{eq:BK}) to momentum space, it can be further
transformed into a form which depends only on logarithmic momentum
variable $\xi \sim \ln k^2$.  In this form the kernel involves only
1-dimensional integral, and it becomes easy to include a very large
range of momenta in the calculation; in our case, more than 20 orders
of magnitude.  Thus, this enables us to obtain a cutoff-independent
continuum limit result for the evolution exponent $\lambda$. The key
result we obtained with this method was already shown in
Fig.~\ref{fig:jimwlk-UV-extrap} for comparison with the JIMWLK
extrapolation. 

\begin{figure}[htbp]
  \centering
\begin{minipage}[t]{7.8cm}
  \includegraphics[width=7.5cm]{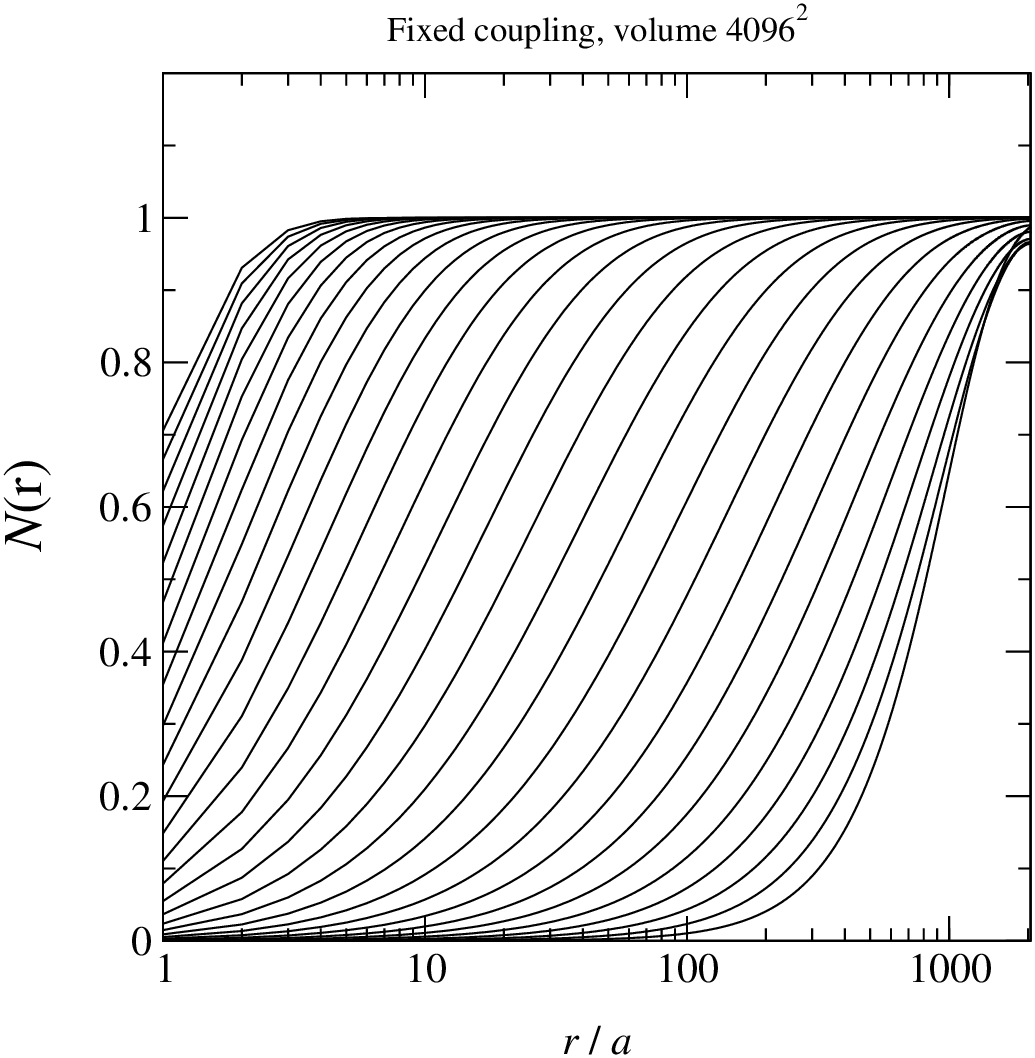}
  \caption{\small The evolution of $N(\br)$ on a $4096^2$ lattice, 
   measured along on-axis direction.  The curves are plotted with
   interval $\delta\tau = 0.25/\alpha$, and the total evolution time
   is $7/\alpha$.}
  \label{fig:N-BK}
\end{minipage}
\hspace{.5cm}
\begin{minipage}[t]{7.8cm}
\includegraphics[width=7.5cm]{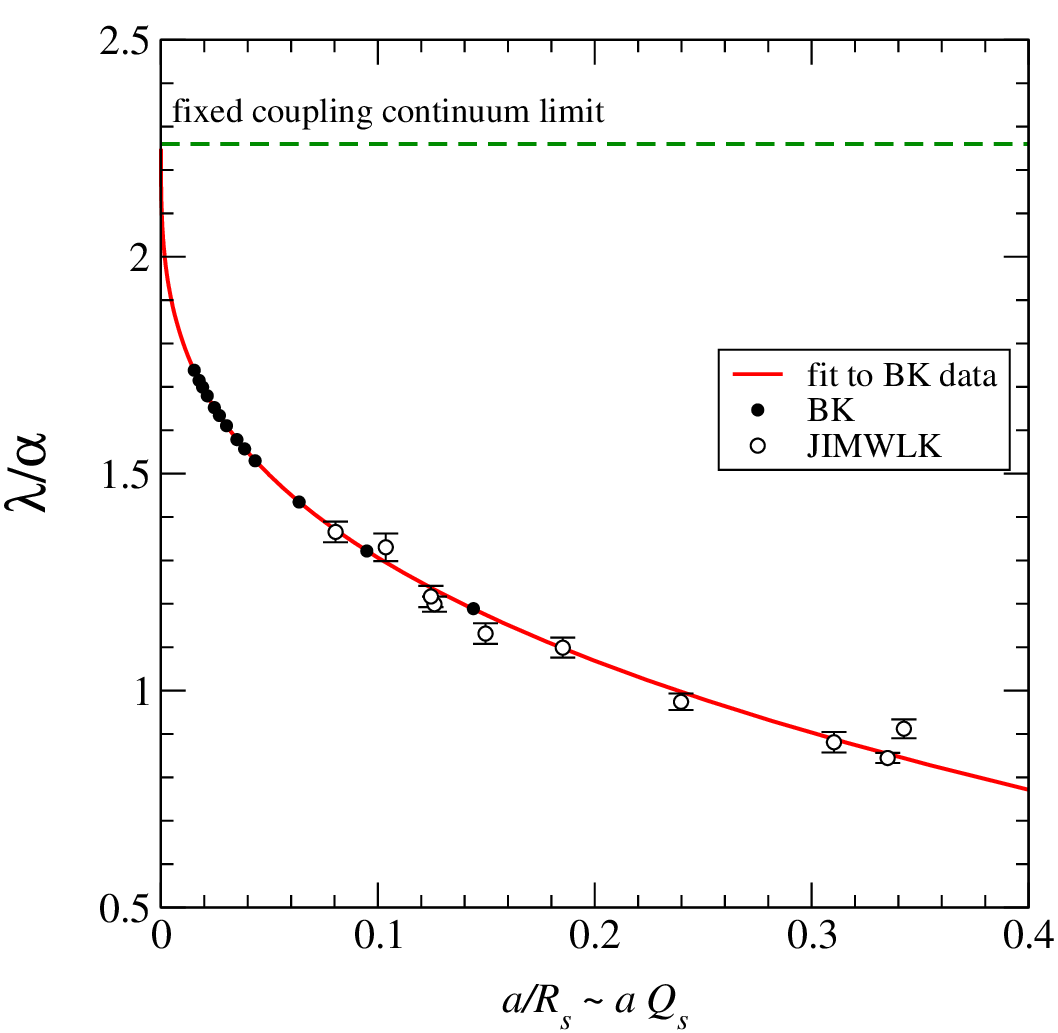}
  \caption{\small Continuum extrapolations for $\lambda$ using JIMWLK and 
    BK simulations. The UV cutoff is plotted in units of the physical
    scale $R_s$. The continuous line is 
   $\lambda/\alpha_s = 2.26 -
    2.0(a/R_s)^{0.32}$ }
  \label{fig:JIMWL+BK-fixed-extrap}
\end{minipage}
\end{figure}

In Fig.~\ref{fig:N-BK} we show the evolution of $N$ measured from
$4096^2$ lattice.  The similarity with the dipole function of JIMWLK
equation is striking, see Fig.~\ref{fig:jimwlk-evo}.  Again we see the
initial settling towards the scaling form, which evolves by
shifting towards smaller $R$ while preserving the shape.  Also here we
see that, even with a scaling form reached, the spacing of curves and
hence $\lambda$ is not constant. This again is due to the growing
influence of the UV cutoff. 

We are now in a position to complement our results from the JIMWLK
equation.  We extract $\lambda$ in the same way as in
Figs.~\ref{fig:lambda-r} and \ref{fig:jimwlk-UV-extrap}.  This time,
there are no statistical errors in the measurements of $\lambda$.  A
function of form $\lambda/\alpha_s = c_1 + c_2 (a/R_s)^\nu$ is an
excellent fit to the data, with result $c_1 = 2.26$, $c_2=2.0$ and
$\nu=0.32$.  The $a\rightarrow 0$ result agrees perfectly with the
momentum space result.  This is shown in
Fig.~\ref{fig:JIMWL+BK-fixed-extrap}.  The agreement of BK and JIMWLK
results at corresponding cutoff scales is remarkable over the whole
range. Most striking is an evaluation of the active phase space. While
on the IR side activity is limited to within one order of magnitude of
the saturation scale $Q_s$, on the UV side we need the cutoff about 6
orders of magnitude larger than $Q_s$ to get within 1\% of the
continuum value.as is easily extracted from the power law fit.

This situation is clearly unnatural: large open phase space of this
magnitude usually generates large logarithmic corrections that need to
be resummed to get physically reliable results. In the following we
will argue that the main source for such corrections are due to
running coupling and study the effects which, among other new {\em
  qualitative} features, will bring a large reduction of active phase
space in the context of the BK equation that erases the potential for
large logarithmic corrections.

\section{Running coupling effects}
\label{sec:runn-coupl-effects}
To account for running coupling effects in the simplest possible way,
we decide to have the coupling run at the natural scale present in the
BK-equation, the size of the parent dipole $(\bm{x}-\bm{y})^2$, i.e.
in Eq.~(\ref{eq:BK}), we replace the constant $\alpha_s$ by the one
loop running coupling at the scale $1/(\bm{x}-\bm{y})^2$:
\begin{equation}
  \label{eq:d-running}
  \alpha_s\to 
  \alpha_s(1/(\bm{x}-\bm{y})^2) = 
  \frac{4\pi}{\beta_0\ln\frac{1}{(\bm{x}-\bm{y})^2
      \Lambda^2
    }} \hspace{1cm}\text{with}\hspace{1cm} \beta_0=(11 N_c -2 N_f)/3
\end{equation}
$\Lambda$ in Eq.~\eqref{eq:d-running} is related to
$\Lambda_{\text{QCD}}$ by a simple factor of the order of $2\pi$.
Such a simple replacement, of course, is not what we expect to arise
from an exact calculation in which the two loop contributions that
generate the running will emerge underneath the integral over
$\bm{z}$. This comes on top of the fact that at next to leading order
there will be corrections beyond the running coupling effects that are
entirely unaccounted for by this substitution. Nevertheless, we would
like to argue that Eq.~(\ref{eq:d-running}), when used in the BK
equation captures the leading effect associated with the
``extraneous'' phase space uncovered in
Sec.~\ref{sec:bk-jimwlk-compared}. This has been the outcome of a long
discussion in the context of the BFKL equation and its next to leading
order corrections~\cite{Fadin:1995xg, Fadin:1996zv, Fadin:1998hr,
  Fadin:1998py, Colferai:1999em, Thorne:1999rb, Thorne:2001nr} Here we
simply take this as our starting point with the attitude that a scale
setting technique like the one employed in Eq.~(\ref{eq:d-running}) is
but the easiest way to get even quantitatively reliable initial
results. We plan on supplementing this with more detailed studies
based on dispersive methods of implementing the running coupling
effects.

With running coupling we have to face the presence of the Landau pole
in Eq.~\eqref{eq:d-running}. The main point here is, that while indeed
we do not control what is happening at that scale we are justified in
providing a prescription for its treatment. With $Q_s \gg
\Lambda_{\text{QCD}}$ the integral on the r.h.s.\ of the BK equation
yields extremely small contributions at scales of the order of
$\Lambda_{\text{QCD}}$, typically of the order of $10^{-7}$ times what
we see around $Q_s$. This is the reason why we are allowed to ignore
whatever happens there and to freeze any correlators such as $N$ at
these scales at the values provided by the initial condition -- a
reasoning that is already necessary at fixed coupling in some sense.
If we turn on running coupling then, the physical thing to do is to
switch off the influence of the divergence at the Landau pole, for
instance by freezing the coupling below some scale $\mu_0$.  Of course
we make sure that the details of where we do that will not affect our
results.

Such an argument about the integrals on the r.h.s.\ is valid at
$b\approx 0$ or infinite transverse target size where the nonlinearity
is present and sizable everywhere we look. For large $b$ one
necessarily enters the region where the nonlinearity is small and
there one would remain sensitive to any value of $\mu_0$. That is our
reason for not considering such situations.

\begin{figure}[htbp]
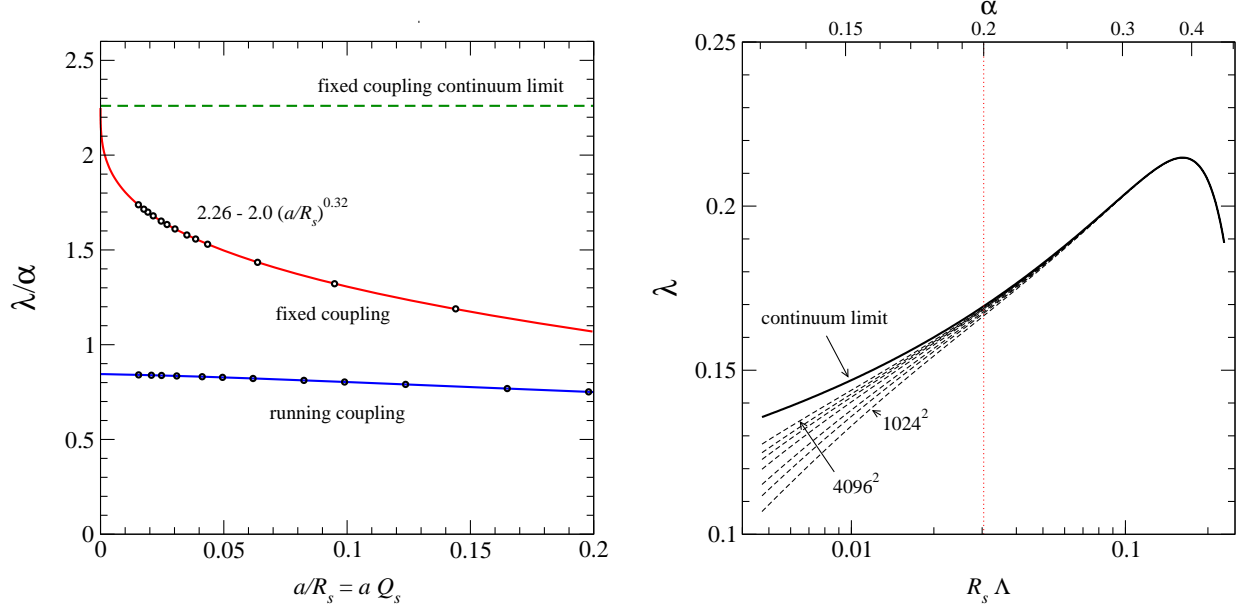

  \centering 
  \begin{minipage}[t]{7.8cm}
\includegraphics[width=8cm]{bk_lambda_fix_running}
  \end{minipage}
  \hspace{.5cm}
  \begin{minipage}[t]{7.8cm}
\includegraphics[width=7.8cm]{bkr2_lambda_vol}
  \end{minipage}  
 \caption{\small Left: example of a continuum extrapolation of 
    $\lambda/\alpha_s(Q_s(\tau)^2)$ for a fixed initial condition
    at a $\tau$-value where $\alpha_s(Q_s(\tau)^2) = 0.2$, compared 
    to the fixed coupling case.  Right: $\lambda(R_s(\tau))$, measured
    using different lattice cutoffs, together with the extrapolated continuum
    limit curve.  The dashed vertical line indicates the point where
    the values on the left are extracted.}
  \label{fig:runn-extrap}
\end{figure}

\begin{figure}[htbp]
  \centering
 \includegraphics[width=7cm]{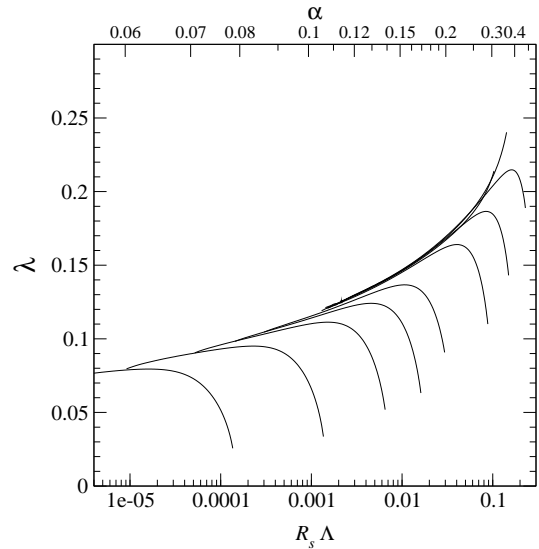}
  \caption{\small $\lambda=\partial_\tau \ln Q_s(\tau) $ for different
    initial conditions.  All curves starting from Gaussian initial
    conditions stay below an asymptotic line along which we find
    (approximately) scaling dipole functions $N$.  }
  \label{fig:runn-lambda}
\end{figure}

Let us now explore how this modification affects evolution.  The
crucial first observation is a clear reduction of active phase space
as shown in Fig.~\ref{fig:runn-extrap}: on the left panel
we show an example for the
extrapolation of $\lambda/\alpha_s(Q_s(\tau)^2)$ to its continuum
value at one value of $\tau$ (corresponding to $\alpha_s(Q_s(\tau)^2) = 0.2$)
and compare to the fixed coupling case.  
With running coupling we are able to reach the continuum limit already
for relatively small lattices, i.e.  with a UV cutoff that would be
orders of magnitude too small in the corresponding fixed coupling run.
Note that as a consequence the continuum value is reduced by a
noticeable factor.  On the right panel we demonstrate how the
extrapolation was done: for a given initial condition we perform runs
with different lattice spacings and measure $\lambda(\tau)$ and
$R_s(\tau)$.  We get a reliable continuum limit extrapolation for
$\lambda$ in a wide range of $\tau$-values; when $R_s$ becomes small
($\sim 5$--$10 a$) the measurements start to fan out.  However, as will be
shown in Fig.~\ref{fig:runn-lambda}, we can extend the measurements
by starting from different initial $R_s$-values.

We will turn to a more systematic study below and
only emphasize that the results given in what follows are physical in the
sense that they apply to the continuum limit.

First and foremost, the running coupling reflects the breaking of
scale invariance. It would be natural to expect that the dipole
function now might never exhibit the scaling behavior of
Eq.~\eqref{eq:BK-scaling} that was so characteristic at large $\tau$
for fixed coupling. In fact numerical evidence suggests otherwise. As
shown in Fig.~\ref{fig:runn-scaling} which shows results
$N_{\tau,\bm{x y}}$ for large $\tau$ against $r\,Q_s(\tau)$. The
different curves only slightly deviate from each other: scaling is
only violated very weakly. We attribute this to the fact that
$Q_s(\tau)\gg \Lambda_{\text{QCD}}$ and thus the scale breaking is not
strongly communicated to the active modes.
\begin{subfigures}
\begin{figure}[htbp]
  \centering
  \begin{minipage}{7.5cm}
  \includegraphics[width=7cm]{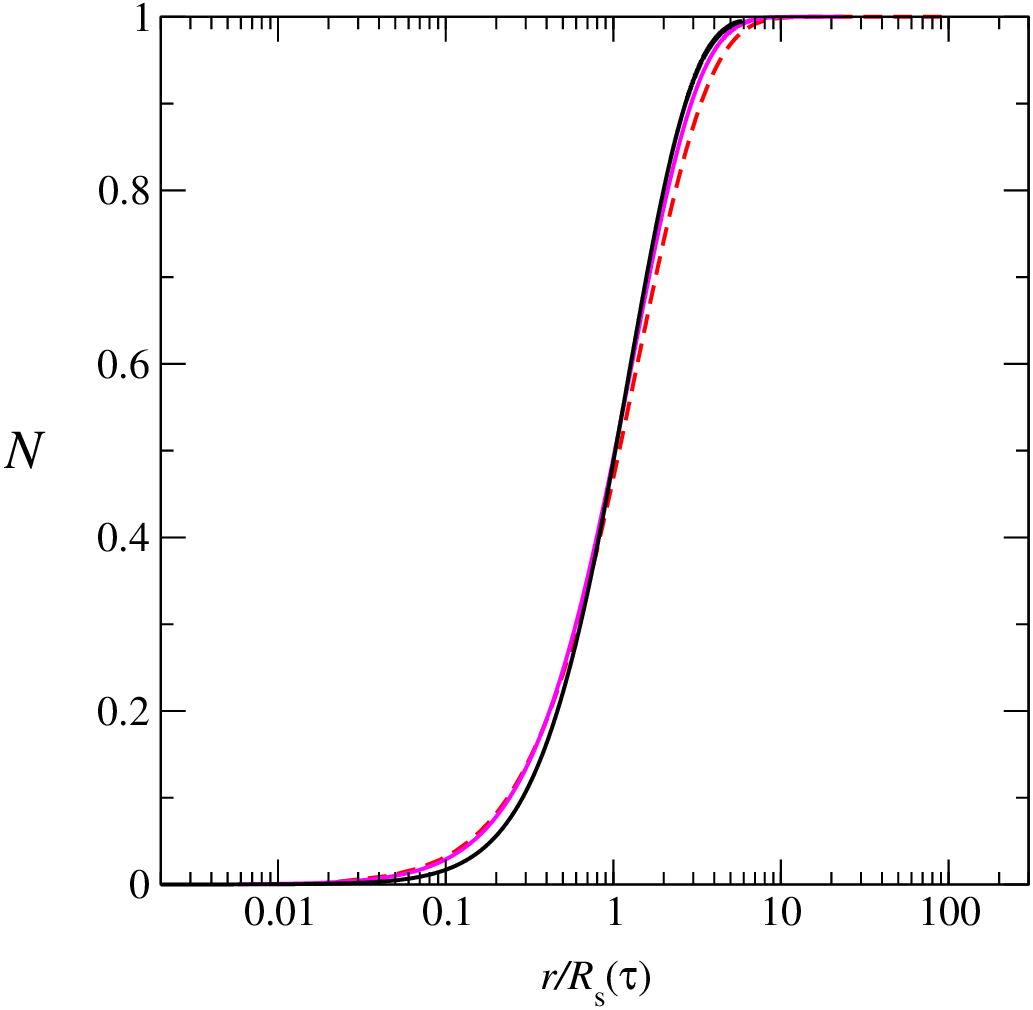}
  \caption{\small Approximate scaling behavior of $N_{\tau,\bm{x y}}$ 
    on the asymptotic line of Fig.~\ref{fig:runn-lambda}.  }
  \label{fig:runn-scaling}
  \end{minipage}  
\hspace{.5cm}
  \begin{minipage}{7.5cm}
   \includegraphics[width=7cm]{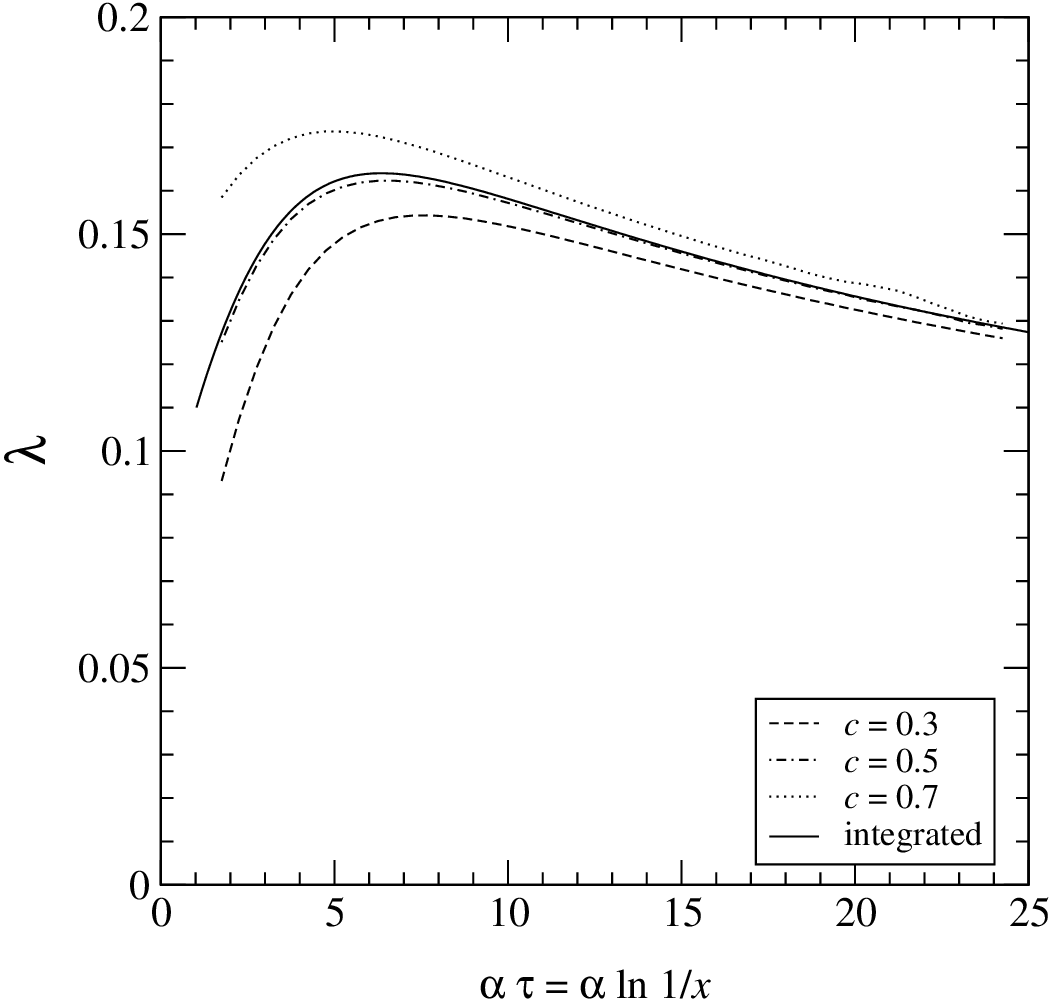}
  \caption{\small Scheme and $\tau$ dependence of $\lambda$ for one 
    initial condition}
  \label{fig:lambda-scheme-tau}
  \end{minipage}
\end{figure}
\end{subfigures}
Where the form of $N_{\tau,\bm{x y}}$ shows approximate scaling it
makes sense to talk about $Q_s$ in complete analogy to the fixed
coupling case. Again defining the rate of change $\lambda$ according to
\begin{equation}
  \label{eq:lambdadef}
  \lambda(\tau):=\partial_\tau \ln Q_s(\tau) \ ,
\end{equation}
leads to an almost unique and scheme independent quantity.  The main
new feature is that now, in contrast to the fixed coupling case,
$\lambda$ is $\tau$-dependent.  Both scheme independence in the near
scaling region and $\tau$-dependence of $\lambda$ are shown for one
initial condition in Fig.~\ref{fig:lambda-scheme-tau}. The curves
correspond to different values of $c$ in Eq.~\eqref{eq:R_s-def-c} and
a direct definition of $\lambda$ via a moment of the evolution
equation to be given below. We see first a strong rise which occurs in
the pre scaling region that turns into a slow decrease in the ``near''
scaling region where the curves converge to a single line. The
accuracy to which the the curves agree in the near scaling region at
large $\tau$ can be used as a measure of the quality of the scaling.

Since in the scaling region, $Q_s(\tau)$ is the only scale left in the
problem, it is natural to plot $\lambda(\tau)$ against $R_s\sim
1/Q_s$.  This allows to compare the asymptotic behavior of simulations
with different initial conditions as shown in
Fig.~\ref{fig:runn-lambda}.  In this plot we find that all $\lambda$
values obtained from Gaussian initial conditions are limited from
above by an asymptotic curve that falls with growing $\tau$ (from left
to right in the plot). It is on this asymptotic curve that the scaling
shown in Fig.~\ref{fig:runn-scaling} holds.

With the global features established let us turn back to a
quantitative analysis of active phase space and a study of the $\tau$
dependence in the near scaling region.

\begin{figure}[htbp]
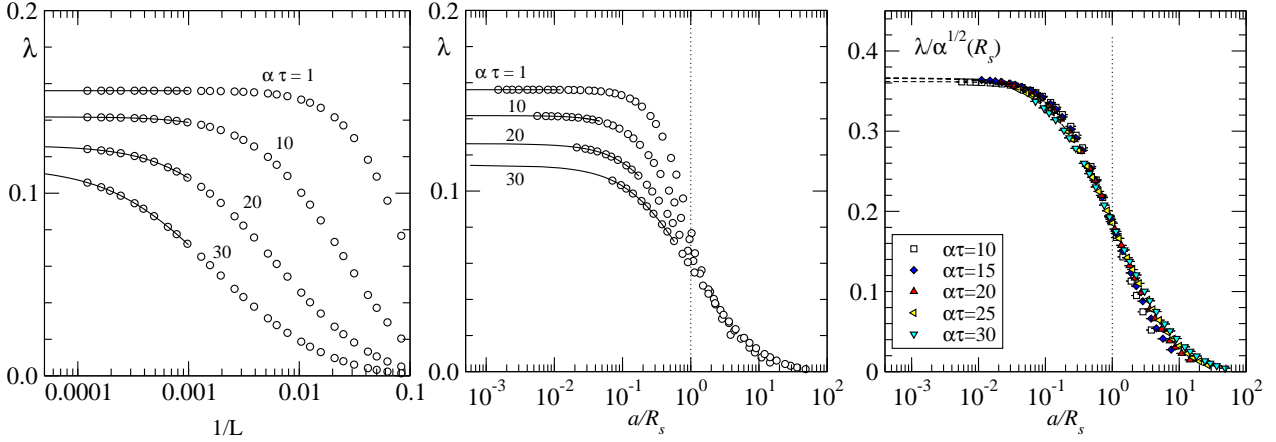

  \centering
\includegraphics[height=5.8cm]{bkr_r2b_lsize2}%
\includegraphics[height=5.8cm]{bkr_r2b_aRsize2}%
\includegraphics[height=5.7cm]{bkr_r2b_aRscaled2}%
\caption{\small UV dependence of the BK-equation on the asymptotic
  line after the parent dipole scale is taken to determine the
  coupling. Left: $\lambda$ against $a$ in units of lattice size.
  Middle: $\lambda$ against $a$ in units of $R_s$, the natural units.
  Right: the same rescaled by $\sqrt{\alpha_s(Q_s^2)}$. Active phase
  space is limited in the UV (and IR) by $5$ to $10\times Q_s$.  The
  $\tau$-dependence of $\lambda$ is, to a good approximation, given by
  $\lambda(\tau)=\lambda_0\sqrt{\alpha_s(Q_s(\tau)^2)}$. }
  \label{fig:runn-phase-space-sqrt-dep}
\end{figure}

To this end we first show a detailed analysis of the continuum
extrapolation of one trajectory at several values in $\tau$ in
Fig.~\ref{fig:runn-phase-space-sqrt-dep}.  On physics grounds we
expect the UV cutoff needed in a given case to be determined by $Q_s$
at that the $\tau$ considered. In
Fig.~\ref{fig:runn-phase-space-sqrt-dep}, left, we see that with
growing $\tau$ the extrapolations of $\lambda$ towards the continuum
limit at the left reach the asymptotic plateau later and later. That
this indeed follows $Q_s$ is shown in the middle.  What remains open
in these plots is the precise form of the $\tau$-dependence of
$\lambda$.  The numerical result here is surprising: our simulations
show a clear scaling of $\lambda/\sqrt{\alpha_s(Q_s(\tau))}$ in the
asymptotic region over the range explored in this simulation. This is
shown in Fig.~\ref{fig:runn-phase-space-sqrt-dep}, right, which
summarizes our results on $\tau$ and scale dependence for $\lambda$ in
the range of $Q_s$ values appearing in this simulation.

Active phase space is clearly centered around $Q_s(\tau)$ both in the
UV and the IR. The reasons for the boundedness are very different on
the two sides: while the IR safety is induced by the nonlinear effects
already present at fixed coupling, the limited range into the UV is
solely caused by the running of the coupling.

While it is natural to express the $\tau$ dependence of $\lambda$ via
$Q_s(\tau)$, it is rather surprising that for the small values of the
coupling constant in the range covered in
Fig.~\ref{fig:runn-phase-space-sqrt-dep} that an expansion of
$\lambda$ of the form
\begin{equation}
  \label{eq:hypothetical}
  \lambda(\tau) = c \alpha_s(Q_s(\tau)^2) + \text{small corrections}
\end{equation}
appears to be totally inadequate and instead a fit of the form
\begin{equation}
  \label{eq:sqrt-lambda}
  \lambda(\tau)=\lambda_0\sqrt{\alpha_s(Q_s(\tau)^2)}
\end{equation}
appears to be surprisingly successful. If the latter provides a good
fit, then the corrections in Eq.~\eqref{eq:hypothetical} must be
important in the range considered. As it turns out, the fact that
$\lambda(\tau)$ can not be expanded in a simple power series in
$\alpha_s(Q_s(\tau)^2)$ is linked with the large reduction in phase
space compared to the fixed coupling case. To see this, let us use an
idea by McLerran, Iancu, and Itakura~\cite{Iancu:2002tr} that was
introduced in the fixed coupling case to show that there $\lambda$ is
necessarily constant in the scaling region.

They assume (in the context of the fixed coupling BK equation) that
$N_{\tau\ \bm{x y}}$ depends only on the combination
$(\bm{x}-\bm{y})^2 Q_s(\tau)^2=:\bm{r}^2 Q_s(\tau)^2 $.  Then
$\tau$-derivatives can be traded for $\bm{r}^2$ derivatives and the
l.h.s.\ of the BK equation is easily rewritten as
\begin{equation}
    \label{eq:BK-lhs-scaling}
\partial_\tau N({\bm{r}^2 Q_s(\tau)^2}) =
  \bm{r}^2 \partial_{\bm{r}^2} N({\bm{r}^2 Q_s(\tau)^2}) \partial_\tau
  \ln Q_s(\tau)^2 = \bm{r}^2 \partial_{\bm{r}^2} N({\bm{r}^2
    Q_s(\tau)^2}) 2 \lambda_\tau \ .    
\end{equation}
Taking into account the spatial boundary conditions imposed by ``color
transparency'', the fact that $N(\bm{r}^2=0)=0$ and
$N(\bm{r}^2=\infty)=1$, leads us to take the zeroth moment of the BK
equation --to integrate with $\frac{d^2r}{\bm{r}^2}$--~in order to
isolate $\lambda$ on the l.h.s.. The r.h.s then provides an integral
expression for it in form of its zeroth moment:
\begin{equation}
    \label{eq:lambda-integral-def}
    2 \pi \lambda(\tau) = \frac{\alpha_s N_c}{2\pi^2} \int 
    \frac{d^2r d^2z }{\bm{u}^2\bm{v^2}}\big( N(\bm{u}^2Q_s^2) 
      + N(\bm{v}^2Q_s^2) - N(\bm{v}^2Q_s^2) 
      - N(\bm{u}^2Q_s^2) N(\bm{v}^2Q_s^2)
      \big)
\end{equation}
where we have set $\bm{u}=\bm{x}-\bm{z}$ $\bm{v}=\bm{z}-\bm{y}$ for
compactness. Due to scale invariance of the integral, the r.h.s.\ is
indeed independent of $\tau$: the scale common to all $N$ can be
chosen at will without changing the integral. Indeed, in the scaling
region Eq.~\eqref{eq:lambda-integral-def} is completely equivalent to
Eq.~\eqref{eq:lambdadef}. Outside that region it can be
used as an alternative definition of a quantity that gives a
qualitative description of the rate of evolution with the same
justification as any other definition of $\lambda$ or $Q_s$. Its main
advantage is that now it is directly related to the integral
expression driving the evolution step and we can directly analyze
active phase space in this expression where needed.

On the asymptotic line with near scaling the argument is only
approximate, but to good accuracy we obtain
\begin{equation}
    \label{eq:running-lambda-integral-def}
    2\pi \lambda(\tau) = \frac{N_c}{2\pi^2} \int 
    \frac{d^2r d^2z }{\bm{u}^2\bm{v^2}}
    \alpha_s(1/\bm{r}^2)
    \big( N(\bm{u}^2Q_s^2) 
      + N(\bm{v}^2Q_s^2) - N(\bm{v}^2Q_s^2) 
      - N(\bm{u}^2Q_s^2) N(\bm{v}^2Q_s^2)
      \big)+\ldots
\end{equation}
and we can take the r.h.s.\ as a definition of $\lambda(\tau)$,
ignoring the small corrections. Fig.~\ref{fig:lambda-scheme-tau} shows
that we indeed agree with the conventionally measured quantity.

Due to the near scaling of $N$ also in the running coupling case
together with all other ingredients of the integral, it is really only
the presence of the running $\alpha_s$ underneath the integral that
distinguishes the integrals in Eq.~\eqref{eq:lambda-integral-def} and
Eq.~\eqref{eq:running-lambda-integral-def}. To obtain the strong
relative reduction in active phase space observed in
Fig.~\ref{fig:runn-phase-space-sqrt-dep} we need $\alpha_s$ to vary
considerably over the range of scales contributing to the integral in
the fixed coupling case.  Wherever this happens, $\lambda(\tau)$ will
be a nontrivial function of $\alpha_s(Q_s(\tau)^2)$. Conversely, at
asymptotically large $Q_s$, where the $\alpha_s(1/\bm{r}^2)$ probed in
Eq.~\eqref{eq:running-lambda-integral-def} turn essentially constant
over a range comparable to the phase space needed in the fixed
coupling case, we should return to a situation as in
Eq.~\eqref{eq:hypothetical}.  The price to pay is the reopening of
phase space to the size encountered already at fixed coupling.

To illustrate this we list active phase space in terms of $a/R$ needed
to find $\lambda$ within a given percentage of its continuum value at
different $\alpha_s(Q_s(\tau)^2)$ in comparison with the fixed
coupling result.
\begin{table}[htbp]
  \centering
  \begin{tabular}[t]{l|l|l|l|l|l}
deviation from  & $\alpha=.2 $           & $\alpha=.15 $         & $\alpha=.12$            & $\alpha=.1$            & fixed  coupling        \\
continuum value & $\lambda/\alpha =.845$ & $\lambda/\alpha =.96$ & $\lambda/\alpha =1.052$ & $\lambda/\alpha =1.14$ &                        \\
& & & & & \\                                                                                                                               
10\%            & $a/R=0.17$             & $a/R=0.15$            & $a/R=0.12$              & $a/R=0.1$              & $a/R=0.0011$           \\
    5 \%        & $a/R=0.1$              & $a/R=0.08$            & $a/R=0.066$             & $a/R=0.05$             & $a/R=0.00013$          \\
    1 \%        & $a/R=0.027$            & $a/R=0.015$           & $a/R=0.007$             & $a/R=0.0012$           & $a/R=8.2\cdot 10^{-7}$ \\
\end{tabular}
\caption{\small Phase space needed to extract continuum values for
    $\lambda$ is reduced for larger values of the coupling. The fixed 
    coupling situation reemerges asymptotically.}
  \label{tab:reopening of phase space}
\end{table}

In the far asymptotic region we are back to the unsatisfactory
situation that we most likely miss important corrections with the
additional drawback that we now have no real idea of what those might
be. Fortunately, this would appear to occur only at such extremely
large values of $Q_s$ for this to be purely academic.

\begin{figure}[htbp]
  \centering
  \includegraphics[width=8cm]{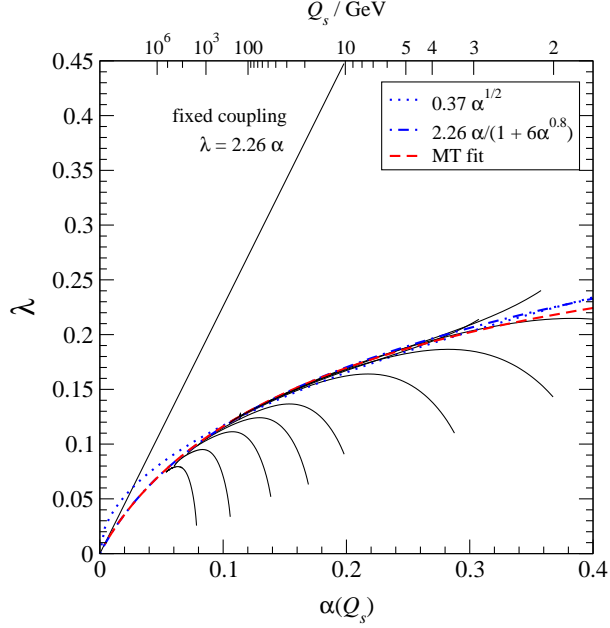}
  \caption{\small
    $\lambda(\tau)$ as a function of $\alpha_s(Q_s(\tau)^2)$. The
    central region of the asymptotic line where the running of the
    coupling reduces phase space is best fitted by a
    $\lambda(\tau)=\lambda_0\sqrt{\alpha_s(Q_s(\tau)^2)}$ or
    equivalently by the use of Eq.~\eqref{eq:dionfit}. At large $Q_s$
    the curve turns towards a form that allows a perturbative
    expansion of the form $\lambda(\tau)=c \alpha_s(Q_s(\tau)^2)
    +\text{small corrections}$ at the price of a reopening of phase
    space.  }
  \label{fig:assesing-sqrt-running}
\end{figure}

With this caveat, we are able to understand the behavior of $\lambda$
on most of the asymptotic line. Fig. \ref{fig:assesing-sqrt-running}
shows the behavior of $\lambda$ as a function of
$\alpha_s(Q_s(\tau)^2)$ and $Q_s$ demonstrating in the center the
large range over which a fit according to Eq.~\eqref{eq:sqrt-lambda}
is successful. To the left of this, we see a deviation that would
appear to be consistent with a turnover towards a linear behavior in
$\alpha_s$. Indeed a purely phenomenological two parameter ansatz of
the form
\begin{equation}
  \label{eq:lambda-alpha-pade-fit}
  \lambda(\tau) = \frac{2.26 \alpha_s(Q_s(\tau)^2)}{%
      1+a \alpha_s(Q_s(\tau)^2)^b}
\end{equation}
provides a fit consistent with all the available information.  Here
$2.26$ is the value obtained from the fixed coupling BK-simulation
which must emerge due our argument above and $a$ and $b$ have been
fitted to be $6$ and $0.8$ respectively.

To establish the link to the asymptotic behavior given
in~\cite{Mueller:2002zm,Triantafyllopoulos:2002nz}, we first observe that the
small coupling limit of Eq.~\eqref{eq:lambda-alpha-pade-fit} with its
leading
\begin{equation}
  \label{eq:running-leading-alpha}
   \lambda(\tau) = 2.26 \alpha_s(Q_s(\tau)^2) +\text{small corrections}
\end{equation}
is in full agreement with the asymptotic behavior derived
in~\cite{Mueller:2002zm,Triantafyllopoulos:2002nz}. This we render
here as
\begin{equation}
  \label{eq:dionlambda}
  \lambda = \frac{.90}{\sqrt{\tau+Y_0}} - \frac{0.47}{(\tau+Y_0)^{5/6}} 
  + \text{higher inverse powers},
\end{equation}
where $Y_0$ stands for constants determined by initial
conditions.\footnote{This form may be obtained from a $Y$ derivative
  of Eq. (83) in~\cite{Mueller:2002zm} or from Eq. (53)
  of~\cite{Triantafyllopoulos:2002nz}. We have converted their $Y$ to
  our $\tau$ and corrected for a factor of $2$ in the definition of
  $\lambda$ } To show agreement with the first term in
Eq.~\eqref{eq:dionlambda} analytically, let us consider a generic
power like dependence of $\lambda$ on $\alpha_s(Q_s(\tau)^2)$ to
incorporate both Eq.~\eqref{eq:running-leading-alpha} valid at
asymptotically small $\alpha_s$ and Eq.~\eqref{eq:sqrt-lambda} which
gives a simple representation for the bulk of scales. We make the
ansatz
\begin{equation}
  \label{eq:lambda-alpha}
 \lambda_\tau := \partial_\tau \ln(Q_s(\tau)/\Lambda_{\text{QCD}})  
 = \lambda_0\ \big[ \alpha_s(Q_s(\tau)^2)\big]^n \hspace{1cm} 
 \alpha_s(Q_s(\tau)^2) := 
 \frac{4\pi}{\beta_0 \ln(Q_s(\tau)^2/ \Lambda_{\text{QCD}})} 
\end{equation}
with $\beta_0=(11 N_c -2 N_f)/3$, and find
\begin{equation}
  \label{eq:Q_s}
  \frac{Q_s(\tau)^2}{\Lambda_{\text{QCD}}^2} = 
  \exp\Big\{ \Big[(n+1) 2 \lambda_0 
  \big(\frac{4\pi}{\beta_0}\big)^n(\tau-\tau_0)
 +\big[\ln(\frac{Q_s(\tau_0)^2}{\Lambda_{\text{QCD}}^2})\big]^{n+1} 
 \Big]^{\frac{1}{n+1}} \Big\} \ .
\end{equation}
Recovering $\lambda$ from this we find
\begin{equation}
  \label{eq:lambda-as-function-of-tau}
    \lambda(\tau) = 
 \frac{ \lambda_0 \sqrt{\frac{4\pi}{\beta_0}}}{\Big[
     (n+1) 2 \lambda_0 \sqrt{\frac{4\pi}{\beta_0}} (\tau-\tau_0)
 +\big[\ln(\frac{Q_s(\tau_0)^2}{\Lambda_{\text{QCD}}^2})\big]^{n+1} 
 \Big]^{\frac{1}{n+1}-1}} 
 \xrightarrow{n\to 1, \lambda_0\to 2.26} 
 \frac{.88}{\sqrt{\tau+Y_0}}
\end{equation}
so that the leading term of Eq.~\eqref{eq:dionlambda} corresponds to
$n=1$ as expected. 

To extend the comparison beyond that we simply integrate $\lambda$ in
Eq.~\eqref{eq:dionlambda} back to get an expression for $Q_s$ and fit
the integration constant $c$ to match our simulations. [Alternatively
we could have directly used the expressions for $Q_s$
in~\cite{Mueller:2002zm,Triantafyllopoulos:2002nz} and adjusted the
unknown parameter.] We find that
\begin{equation}
  \label{eq:dionfit}
  \frac{Q_s(\tau)}{\Lambda_{\text{QCD}}} = c \cdot 
  \exp[ 2\cdot 0.9 (\tau+Y_0)^{\frac{1}{2}} 
  -6\cdot 0.47 (\tau+Y_0)^{\frac{1}{6}}]
\end{equation}
with $c=1.5$ leads to excellent agreement over the whole range. This
is shown in Fig.~\ref{fig:assesing-sqrt-running}.

To summarize the results of this section we state that
Fig.~\ref{fig:runn-phase-space-sqrt-dep} also shows impressively that
the contributions to $\lambda$ arise from within little more than an
order of magnitude of the saturation scale [the region in cutoffs over
which there is an appreciable change in the value extracted]. This
indicates that the dominant phase space corrections are indeed taken
into account and the basic physics ideas that had initially motivated
the density resummations are realized. While the IR side of the
resummation was working already at fixed coupling, the UV side of
phase space is only tamed if the coupling runs. Only with both the
nonlinearities and a running coupling are we justified to claim that
the evolution is dominated by distances of the order of $R_s$ and thus
the coupling involved in typical radiation events during evolution is
of order $\alpha_s(Q_S(\tau)^2)$. Fig.~\ref{fig:assesing-sqrt-running}
shows that on the asymptotic line
$\lambda(\tau)=\lambda_0\sqrt{\alpha_s(Q_s(\tau)^2)}$ over most of the
region where running coupling limits phase space.

\section{Evolution and $A$-dependence}
\label{sec:evol-a-depend}

It has been argued in the context of the
McLerran-Venugopalan model that going to larger nuclear targets
(larger nuclear number $A$) ought to have an enhancing effect on the
importance of the nonlinearities encountered. If we are at small
enough $x$ (high enough energies) for the projectile to punch straight
through the target going to a larger target means interaction with
more scattering centers as indicated in
Fig.~\ref{fig:small-x-dis-geom-rest}. Taking these to be color charges
located in individual nucleons inside the target nucleus the
interaction will occur with more and more uncorrelated color charges
the larger the target. Viewed from up front this amounts to a smaller
correlation length in the transverse direction. In terms of $Q_s$ this
would lead to
\begin{equation}
  \label{eq:QsptoQsA}
  Q_s^A(\tau)^2 = A^{\frac{1}{3}}  Q_s^p(\tau)^2
\end{equation}
without a clear delineation on the $\tau$ values for which this would
be appropriate. This simple argument has been uniformly used
 in~\cite{Freund:2002ux} to relate cross section- (or $F_2$-)
measurements on protons at HERA with measurements on nuclei at NMC and
E665. Despite the fact that the nuclear data used are at rather small
$Q^2$ and large $x$ the agreement found was surprisingly good.

In the context of small $x$ evolution such a statement becomes less
trivial: the whole idea of deriving a small $x$ evolution equation is
based on premise that the logarithmic corrections determining the
evolution step are entirely target independent. The only place where
target dependence is allowed to enter is the initial condition.

In simple cases, these two notions are not conflicting with each
other: indeed a rescaling like in Eq.~\eqref{eq:QsptoQsA} is fully
compatible with fixed coupling evolution on the asymptotic line, where
(in the continuum limit)
\begin{equation}
  \label{eq:fixed-A-scaling}
  Q_s(\tau)= e^{\lambda\tau} Q_s(\tau_0)
\end{equation}
with a universal value for $\lambda$, so that a simple rescaling of
the properties of the initial condition according to
\begin{equation}
  \label{eq:initial-A-scaling}
  Q_s^A(\tau_0)^2 = A^{\frac{1}{3}}  Q_s^p(\tau_0)^2 \ 
\end{equation}
automatically yields Eq.~\eqref{eq:QsptoQsA} at all $\tau$ declaring
it fully compatible with evolution. 

At running coupling, however this ceases to be the case. The
straightforward physical argument is that evolution is characterized by
$\alpha_s(Q_s^2)$ which initially for protons is larger than for
nuclei because of the difference in saturation scales at the initial
condition. In this situation evolution will be faster for smaller
targets and eventually catch up with that of the larger ones.

Technically we find that with running coupling (again restricting
ourselves to the asymptotic line), we find a $\tau$ dependent
$\lambda$ that we may parametrize through some function $f$ of
$\alpha_s(Q_s(\tau)^2)$ as as in the previous section:
\begin{equation}
  \label{eq:lambda-of-f-alpha}
  \lambda(\tau)=\partial_\tau Q_s(\tau) = f(\alpha_s(Q_s(\tau)^2))
\end{equation}
In this generic case
\begin{equation}
  \label{eq:Qs-lambda-generic}
   Q_s(\tau)= e^{\int^\tau_{\tau_0} d\tau'  
     f(\alpha_s(Q_s(\tau)^2)} Q_(\tau_0)
\end{equation}
and a rescaling of $Q_s^p(\tau_0)^2$ will not simply factor out as in
Eq.~\eqref{eq:fixed-A-scaling}. Therefore and a simple rescaling like
Eq.~\eqref{eq:QsptoQsA} will conflict with evolution. All
phenomenological fit functions encountered in the previous section
share this property which is at the heart of the reasoning used by
Mueller in~\cite{Mueller:2003bz} to argue that evolution will
eventually erase $A$-dependence in $Q_s$.

To illustrate the point, let trace the the memory loss with a simple
parametrization like Eq.~\eqref{eq:lambda-alpha}, equivalent to a
$\tau$ dependence of $Q_s$ according to Eq.~\eqref{eq:Q_s}.

Scaling in the $A$-dependence at $\tau_0$ according to
$
  Q_s^A(\tau_0)^2 = A^{\frac{1}{3}}  Q_s^p(\tau_0)^p \ 
$
within Eq.~\eqref{eq:Q_s}
 then predicts both $\tau$ and $A$
dependence of the saturation scale.

\begin{figure}[htbp]
  \centering
  \includegraphics[width=8cm]{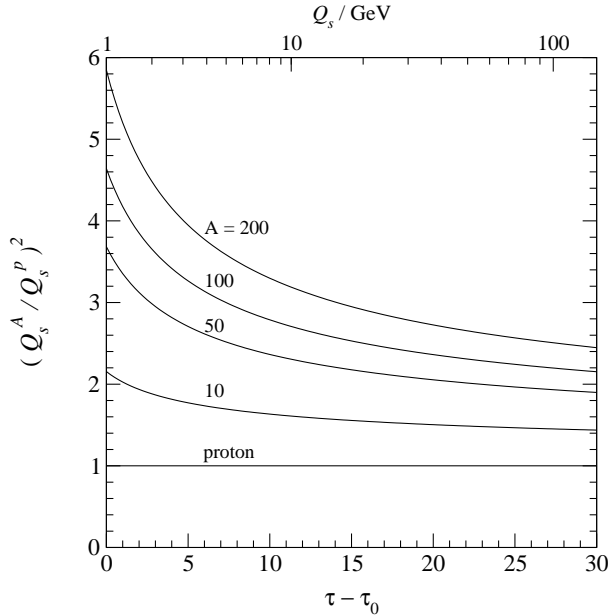}
  \caption{\small $A$-dependence for proton saturation scales from 
    1 to 100 GeV. }
  \label{fig:A-dep}
\end{figure}

We find that evolution will indeed strive to wash out $A$-dependence
in qualitative agreement with~\cite{Mueller:2003bz}.  Indeed, the
ratio of saturation scales in the bulk of the region shown in
Fig.~\ref{fig:assesing-sqrt-running} then behaves like
\begin{equation}
  \label{eq:Q_s-scaling}
  \frac{Q^A_s(\tau)^2}{Q^p_s(\tau)^2} = \frac{
    \exp{ \Big[(n+1) 2\lambda_0 
  \big(\frac{4\pi}{\beta_0}\big)^{n}(\tau-\tau_0)
 +\big[\ln(\frac{A^\frac{1}{3} 
   Q_s(\tau_0)^2}{\Lambda_{\text{QCD}}^2})\big]^{n+1} 
 \Big]^{\frac{1}{n+1}}
}}{  
    \exp{ \Big[(n+1) 2\lambda_0 
  \big(\frac{4\pi}{\beta_0}\big)^{n}(\tau-\tau_0)
 +\big[\ln(\frac{ 
   Q_s(\tau_0)^2}{\Lambda_{\text{QCD}}^2})\big]^{n+1} 
 \Big]^{\frac{1}{n+1}}
}
} \xrightarrow{\tau\to\infty} 1
\ .
\end{equation}
While the parametrization with $n=1/2$ is not applicable at arbitrary
large $\tau$, this formula nevertheless gives a reasonable estimate of
the rate of convergence within its window of validity. This is plotted
in Fig.~\ref{fig:A-dep}. Clearly larger nuclei suffer stronger
``erasing.'' It should be kept in mind that realistically one should
not expect real world experiments to cover more than a few orders of
magnitude in $x$ and hence too large an interval in $\tau-\tau_0$.

The fact that Eq.~\eqref{eq:sqrt-lambda} and the ensuing $\tau$
dependence is only valid on the asymptotic line of
Fig.~\ref{fig:runn-lambda} prevents us from making a direct statement
about current and future experiments. Such stronger statements require
a more thorough study of initial conditions in direct comparison with
existing experiments. If dependence on initial conditions can be
clarified $A$ dependence can be studied numerically also in the
pre-asymptotic region.

\section{Conclusions}
\label{sec:conclusions}

In this paper we have, for the first time, shown results from a direct
numerical implementation of the JIMWLK equations. The simulations have
been performed using a Langevin formulation equivalent to the original
equations. These then have been implemented numerically using a
lattice discretization of transverse space.

Our simulations have been mainly concerned with extracting universal
features and asymptotic behavior induced by the equation itself
without direct comparisons with data and systematic studies of initial
conditions.

We have clearly established infrared stability of the solutions: once
provided with an initial condition that furnishes a (density induced)
saturation scale, or correlation length $R_s\sim 1/Q_s$, the solutions
are completely insensitive to IR cutoffs sufficiently below that
scale. Such behavior had already been found in the large $N_c$ limit
in the context of the BK equation and here merely serves as a
consistency check.

Simulations of the BK equation have up to now been only carried out in
momentum space, our simulations are the first in coordinate space and
thus we are the first to show the approach towards the fixed point at
infinite energy with vanishing correlation length directly.

Beyond confirming analytical results already found
in~\cite{Weigert:2000gi}, we also see the emergence of scaling
solutions, extending earlier results from simulations of the BK
equation to finite $N_c$. Within the class of initial conditions
studied these emerge universally already at finite energy.

The initial conditions studied were all transversally homogeneous and
did not contain large intrinsic $N_c$ violations. We found that JIMWLK
evolution did not enhance these to above the percent level. In such a
situation the BK equation should provide a good approximation.  One
should be cautious to generalize this too far: any type of lumpiness
in the initial condition, hot spots of any kind, would presumably
spoil this feature and the factorization limit would not be adequate.

Simulations of the BK equation had already clearly shown that
evolution populates a large range in phase space of tens of orders of
magnitude, a fact easily attributable to the random walk property of
the BFKL equation which due to density effect is tamed in the IR, but
not in the UV. While previous studies have taken this as a given, we
were forced by the necessity to work on a coordinate lattice to
carefully study the UV properties to obtain a valid continuum
extrapolation.

Already within the JIMWLK simulation we again found evidence for
enormous active phase space that induced relatively large errors in
the continuum extrapolations. Since our initial condition did lead to
simulations which respect factorization we were in a position to refine
this study by implementing the BK-equation in coordinate space with
the same type of lattice regulator used in the JIMWLK simulations.  We
found excellent agreement between the JIMWLK and BK simulations on
finite lattices and were able to confirm that the (constant) evolution
rate of the saturation scale on the asymptotic line $\lambda =
\partial_\tau Q_s(\tau)$ gets contributions from the far UV, up to 6
orders of magnitude away. In such a situation one typically expects
large logarithmic corrections that need to be resummed. Indeed, from
the experience with the BFKL equation at two loop, there was an
obvious candidate for the main effect, the running of the coupling.

This leads us to the second part of the paper. Here we studied the
effect of a naive implementation of running coupling on the same type
of questions as studied above. This has been done in the context of
the BK-equations for reasons both of efficiency and principle. The
latter will be briefly discussed below.

The main effect of running coupling was to drastically improve the
convergence towards the continuum limit over a gigantic range of $Q_s$
values between $1$ and well above $100$GeV.  [Note that the
Golec-Biernat+W\"usthoff fits to HERA data have $Q_s$ rise from $1$ to
$2$GeV from $x=10^{-2}$ to $10^{-4}$.]

Despite the explicit breaking of scale invariance we find that the
dipole function still converges towards a scaling form with all the
$\tau$ dependence taken by the saturation scale $Q_s(\tau)$.

The evolution rate is effected more strongly: $\lambda(\tau) =
\partial_\tau \ln Q_s(\tau)$ now is necessarily $\tau$ dependent.
Within the window of strongly reduced phase space between $1$ and
$100$GeV it is well described by a simple fit of the form
$\lambda(\tau) = .37 \sqrt{\alpha_s(Q_s(\tau)^2)}$. This is strongly
different from the naive expectation that one should find
$\lambda(\tau) = c \alpha_s(Q_s(\tau)^2)+\text{small corrections}$. We
have numerical and analytical indications that such a behavior would
reemerge at extremely large $\tau$ at the price of a reopening of
phase space. The explanation for this is simple: as shown in
Sec.~\ref{sec:runn-coupl-effects}, the running coupling appears in an
integral expression for $\lambda$ that reduces to the corresponding
fixed coupling expression where the running is weak over phase space
active in the integral. Thus a strong reduction in phase space
necessarily means that $\alpha_s$ can not be factored out of the
integral. At extremely large $Q_s$, however, the running will slow
down sufficiently to allow the fixed coupling features to reemerge.
The agreement with~\cite{Mueller:2002zm,Triantafyllopoulos:2002nz} on
the other hand is excellent.

The running coupling induced $\tau$ dependence of $\lambda$ is
responsible for another feature of small $x$ evolution already
discussed by Mueller~\cite{Mueller:2003bz}, the washing out of the $A$
dependence of $Q_s$ inherent to the McLerran-Venugopalan model.
Scaling $Q_s$ at the initial condition at small $\tau_0$ according to
the McLerran-Venugopalan rule $Q_s^A(\tau_0)^2 = A^{\frac{1}{3}}
Q_s^p(\tau_0)^2$ before evolution adds gluons in an obviously $A$
independent manner leads to a fully determined $\tau$ and $A$ behavior
of $Q_s$ in which, because of the smaller $\alpha_s$ at larger scales
evolution of smaller targets can catch up with that of larger ones.
Our simulations give quantitative results on the asymptotic line.
Quantitative studies with specific initial conditions are possible.

The studies presented in this paper have shown the feasibility of
JIMWLK simulations but at the same time have highlighted the necessity
of implementing running coupling corrections to obtain reliable
results. In particular the latter has been done with a naive scale
setting for the running coupling at the parent dipole size. Here
clearly improvements are necessary. A more refined treatment would
most certainly lead to modifications of our quantitative results such
as the fits obtained for the $\tau$-dependence of $\lambda$, although
we expect our qualitative conclusions to remain valid. An example that
clearly demonstrates the insufficiently of the scale setting used is
the fact that such a treatment would not be possible is that such a
scale setting is impossible in the full JIMWLK context. There the
closest thing to a parent dipole running would be to set the scale at
$1/(\bm{x}-\bm{y})^2$ in Eq. \eqref{eq:JIMWLK}. This, however, is
incompatible with BK parent dipole running as is would simply set the
scale to infinity in the $\sigma$ terms of Eq.~\eqref{RG-new-standard}
-- a step incompatible with the derivation of the BK-equation.
Preliminary studies together with Einan Gardi and Andreas Freund have
shown that consistency in this respect can be achieved if one
implements running coupling with dispersive methods (as widely used
for other reasons in the renormalon context), but further work is
necessary to gain thorough understanding of all the issues involved.

Complementary issues like the proper choice of initial conditions
and questions of $b$-dependence need to be studied to be able to put
comparison with experiment on a sound basis.

\paragraph{Acknowledgements:} 
This work as been carried out over a long timespan and we wish to
thank many people for discussions and input at different stages: Yuri
Kovchegov, Einan Gardi, Andreas Freund, Genja Levin, Alex Kovner, and
Larry McLerran. In particular, we wish to thank Al Mueller and
Dionysis Triantafyllopoulos for discussions about the running coupling
results.  During part of this work HW was supported by the DFG
Habilitanden program.

\appendix

\section{From Fokker-Planck to Langevin formulations, the principle}
\label{sec:from-fokker-planck}


Although this topic is covered in many textbooks, we find that the
presentation is often somewhat oldfashioned and the intimate
connection with path-integrals is not always mentioned.  Let us
therefore begin by demonstrating the relationship of Fokker-Planck and
Langevin formulations with the aid of a simple toy example, a particle
diffusion problem, that, for ease of comparison, shares some of the
features of our problem. The toy model equation is
\begin{equation}
  \label{eq:toyFP}
  \partial_\tau P_\tau(\bm{x}) = - H_{\text{FP}}  \ P_\tau(\bm{x})
\end{equation}
with a Fokker-Planck Hamiltonian
\begin{equation}
  \label{eq:toyFPHam}
  H_{\text{FP}} = \frac{1}{2} i\partial_{\bm{x}}^\mu\, 
  \chi_{\mu\nu}(\bm{x})\, i\partial_{\bm{x}}^\nu \ .
\end{equation}
The particles coordinate $\bm{x}$ in $D$ dimensional space corresponds
to the field variables of the original problem and the probability
distribution $P_\tau$ to $\Hat Z_\tau$. It serves to define
correlation function of some operator $O(\bm{x})$ via $\langle
O(\bm{x}) \rangle_\tau = \int d^Dx O(\bm{x}) P_\tau(\bm{x})$

Eq.~(\ref{eq:toyFP}) admits a path-integral solution which is, as
usual, built up from infinitesimal steps in $\tau$: this is a solution
$P(\tau',\bm{x}';\tau,\bm{x})$ to Eq.~\eqref{eq:toyP}, with initial
condition
\begin{equation}
  \label{eq:toyinitcond}
  P(\tau,\bm{x}';\tau,\bm{x}) = \delta^{(D)}(\bm{x}'-\bm{x})
\end{equation}
that is composed of many, infinitesimally small steps at times
$\tau_k:=\tau+k\epsilon$. In this solution,  the product rule
\begin{equation}
  \label{eq:pathint-sol0}
   P(\tau',\bm{x}';\tau,\bm{x}) =\int d^Dx_{n-1} \ldots d^Dx_1\ 
   P(\tau',\bm{x}';\tau_{n-1},\bm{x}) P(\tau_{n-1},\bm{x}_{n-1};\tau_{n-2},\bm{x}_{n-2})\ldots P(\tau_{1},\bm{x}_{1};\tau,\bm{x})
\end{equation} 
expresses the finite $\tau'-\tau$ solution
$P(\tau',\bm{x}';\tau,\bm{x})$ in terms of infinitesimal steps
$P(\tau_i,\bm{x}_i;\tau_{i-1},\bm{x}_{i -1})$. Solution with
arbitrary initial conditions $ P_{\tau_0}(\bm{x})$ are then recovered
via $P_\tau(\bm{x})= \int d^Dy P(\tau,\bm{x};\tau_0,\bm{y})
P_{\tau_0}(\bm{y})$.

The derivation of the infinitesimal step from $\tau_{i-1}$ to
$\tau_i=\tau_{i-1}+\epsilon$ is textbook material and we omit the
step. It reads (the index $i$ on the coordinate also refers to the
time step, not to the vector component)
\begin{equation}
  \label{eq:toyP}
  P(\tau_i,\bm{x}_i;\tau_{i-1},\bm{x}_{i -1})= 
  N \int\!\!d^Dp_i \ e^{-\epsilon \big( \frac{1}{2}
    \bm{p}_i^\alpha \bm{p}_i^\beta \chi^{\alpha \beta}(\bm{x}_{i-1}) 
    + i \bm{p}_i\cdot 
    \big(\frac{\bm{x}_i-\bm{x}_{i-1}}{\epsilon}-\sigma(\bm{x}_{i-1} )\big)
    \big)}+{\cal O}(\epsilon^2)
\end{equation}
where $\sigma_\mu(\bm{x}) := \frac{1}{2} i\partial_{\bm{x}}^\alpha
\chi_{\alpha\mu}(\bm{x})$ and $N$ a coordinate independent
normalization factor.

From here, the step to the Langevin formulation is trivial: as this
expression is quadratic in the momenta we can trivially rewrite it
with the help of an auxiliary variable $\bm{\xi_i}$ [$i$ again the
step label] as
\begin{equation}
  \label{eq:toyPlang0}
   P(\tau_i,\bm{x}_i;\tau_{i-1},\bm{x}_{i-1})= 
   N \int\!\! d^D\xi_i\ \sqrt{\det(\chi(\bm{x}_i))}\
   e^{-\frac{1}{2}\bm{\xi}_i\chi^{-1}(\bm{x}_i)\bm{\xi}_i}\ 
   \delta^{D}\big(\bm{x}_i-[\bm{x}_{i-1}
   +\epsilon(\bm{\xi}_i
   +\sigma(\bm{x}_{i-1})]\big)\ .
\end{equation}
The $\delta^{D}(\ldots)$ arises from the momentum integration and
determines $\bm{x}_i$ in terms of $\bm{x}_{i-1}$ and the correlated
noise $\bm{\xi}_i$.\footnote{Re-expressing the delta function via a
  momentum integral and performing the Gaussian integral over
  $\bm{\xi}_i$ immediately recovers Eq.~\eqref{eq:toyP}} The equation
for $\bm{x}_i$,
\begin{equation}
  \label{eq:toyLang1}
  \bm{x}_i=\bm{x}_{i-1}+\epsilon(\bm{\xi}_i+\sigma(\bm{x}_{i-1})\ ,
\end{equation}
is called the Langevin equation. It contains both a deterministic term
[the $\sigma(\bm{x}_{i-1})$ term] and a stochastic term [the
$\bm{\xi}$ term]. To fully define the problem without any reference to
its path-integral nature one then needs to state the Gaussian nature
of the (correlated) noise separately. This is often done in textbooks.
We find that the path-integral version is much more
elucidating.\footnote{One of the main sources of
  confusion with stochastic differential equations, the choice of
  discretization [the details of the $x_i$ and $x_{i-1}$ dependence]
  and its impact on the form of the equation itself finds its natural
  explanation there. It corresponds to the choice of discretization in
  the path integral solution Eq.~(\ref{eq:toyP}). Our choice here is
  called the Stratonovich form. We might have as well given the Ito
  form by choosing a different discretization for Eq.~(\ref{eq:toyP})
  or yet another version with the same physics content.}

If, as in our case, $\chi$ factorizes as
\begin{equation}
  \label{eq:toyfac}
  \chi_{\mu\nu}(\bm{x})={\cal E}_{\mu a}(\bm{x}){\cal E}_{\nu a}(\bm{x})\ ,
\end{equation}
a simple redefinition of the noise variable leads to a version of the
Langevin formulation with a Gaussian {\em white} noise:
\begin{equation}
  \label{eq:toyPlang1}
   P(\tau_i,\bm{x}_i;\tau_{i-1},\bm{x}_{i-1})= N \int\!\! d^n\xi_i\ 
   e^{-\frac{1}{2}\bm{\xi}_i^2}\ 
    \delta^{D}\big(\bm{x}_i-[\bm{x}_{i-1}+\epsilon({\cal E}(\bm{x}_{i-1})\bm{\xi}_i+\sigma(\bm{x}_{i-1})]\big) \ .
\end{equation}
The correlation is now absorbed into the Langevin
equation\footnote{The dimensions $D$ and $n$ of configuration space
  and noise now need not be equal as is illustrated by our physics
  example. Again,  write the $\delta$-function as a
  momentum integral and carry out the Gaussian integral in
  $\bm{\xi}$ to recover Eq.~\eqref{eq:toyP} from
  Eq.~\eqref{eq:toyPlang1}.} which reads
\begin{equation}
  \label{eq:toyLang2}
  \bm{x}_i=\bm{x}_{i-1}+\epsilon({\cal E}(\bm{x}_{i-1})\bm{\xi}_i+\sigma(\bm{x}_{i-1})\ , 
\end{equation}
Wherever this form is available, it will be the most efficient
version to use in a numerical simulation as the associated white noise
can be generated more efficiently then the correlated noise of the
more general case.

What is left is to sketch a numerical procedure to implement the
solution given in Eqns.~\eqref{eq:pathint-sol0},~\eqref{eq:toyP}
using the expressions Eq.~\eqref{eq:toyPlang0} or
Eq.~\eqref{eq:toyPlang1}. As a first step we replace the average with
the probability distribution $P_\tau(\bm{x})$ by an ensemble average
\begin{equation}
  \label{eq:toy-ensemble-average}
  \langle O(\bm{x}) \rangle_\tau = \int\! d^Dx\  O(\bm{x})\  P_\tau(\bm{x}) 
  \approx 
  \sum\limits_{U\in {\sf E}[ P_\tau]}  O(\bm{x}) 
\end{equation}
where the ensemble is taken according to the probability distribution
$P_\tau(\bm{x})$. This we can easily do for the initial condition at
$\tau_0$. The Langevin equation then allows us to propagate each
ensemble member down in $\tau$ by $\epsilon$, thereby giving us a new
ensemble at $\tau_1=\tau+\epsilon$. Iteration then gives us a chain of
ensembles for a discrete set of $\tau_i$ that provide an approximation
to a set of $P_{\tau_i}(\bm{x})$ in that they allow us to measure any
correlator $\langle O(\bm{x}) \rangle_\tau$ according to
Eq.~\eqref{eq:toy-ensemble-average}.

The above clearly shows the nature of ``Langevin system'' as a way of
writing a path-integral solution to a differential equation that is
particularly easy to implement. Many often confusing issues, such as
the time step discretization issues often discussed, find their
natural resolution in this setting.


\begin{thebibliography}{10}

\bibitem{Mueller:1986wy}
A.~H. Mueller and J.-w. Qiu, {\it Gluon recombination and shadowing at small
  values of x},  {\em Nucl. Phys.} {\bf B268} (1986) 427.

\bibitem{Mueller:1994rr}
A.~H. Mueller, {\it Soft gluons in the infinite momentum wave function and the
  {BFKL} pomeron},  {\em Nucl. Phys.} {\bf B415} (1994) 373--385.

\bibitem{Mueller:1994jq}
A.~H. Mueller and B.~Patel, {\it Single and double {BFKL} pomeron exchange and
  a dipole picture of high-energy hard processes},  {\em Nucl. Phys.} {\bf
  B425} (1994) 471--488, [\href{http://xxx.lanl.gov/abs/hep-ph/9403256}{{\tt
  hep-ph/9403256}}].

\bibitem{McLerran:1994ni}
L.~D. McLerran and R.~Venugopalan, {\it Computing quark and gluon distribution
  functions for very large nuclei},  {\em Phys. Rev.} {\bf D49} (1994)
  2233--2241, [\href{http://xxx.lanl.gov/abs/hep-ph/9309289}{{\tt
  hep-ph/9309289}}].

\bibitem{McLerran:1994ka}
L.~D. McLerran and R.~Venugopalan, {\it Gluon distribution functions for very
  large nuclei at small transverse momentum},  {\em Phys. Rev.} {\bf D49}
  (1994) 3352--3355, [\href{http://xxx.lanl.gov/abs/hep-ph/9311205}{{\tt
  hep-ph/9311205}}].

\bibitem{McLerran:1994vd}
L.~D. McLerran and R.~Venugopalan, {\it Green's functions in the color field of
  a large nucleus},  {\em Phys. Rev.} {\bf D50} (1994) 2225--2233,
  [\href{http://xxx.lanl.gov/abs/hep-ph/9402335}{{\tt hep-ph/9402335}}].

\bibitem{Ayala:1995kg}
A.~Ayala, J.~Jalilian-Marian, L.~D. McLerran, and R.~Venugopalan, {\it The
  gluon propagator in nonabelian {Weizsacker-Williams} fields},  {\em Phys.
  Rev.} {\bf D52} (1995) 2935--2943,
  [\href{http://xxx.lanl.gov/abs/hep-ph/9501324}{{\tt hep-ph/9501324}}].

\bibitem{Kovner:1995ja}
A.~Kovner, L.~D. McLerran, and H.~Weigert, {\it Gluon production from
  non{A}belian {W}eizsacker-{W}illiams fields in nucleus-nucleus collisions},
  {\em Phys. Rev.} {\bf D52} (1995) 6231--6237,
  [\href{http://xxx.lanl.gov/abs/hep-ph/9502289}{{\tt hep-ph/9502289}}].

\bibitem{Ayala:1996hx}
A.~Ayala, J.~Jalilian-Marian, L.~D. McLerran, and R.~Venugopalan, {\it Quantum
  corrections to the {Weizsacker-Williams} gluon distribution function at small
  x},  {\em Phys. Rev.} {\bf D53} (1996) 458--475,
  [\href{http://xxx.lanl.gov/abs/hep-ph/9508302}{{\tt hep-ph/9508302}}].

\bibitem{Kovchegov:1996ty}
Y.~V. Kovchegov, {\it Non-abelian {Weizsaecker-Williams} field and a two-
  dimensional effective color charge density for a very large nucleus},  {\em
  Phys. Rev.} {\bf D54} (1996) 5463--5469,
  [\href{http://xxx.lanl.gov/abs/hep-ph/9605446}{{\tt hep-ph/9605446}}].

\bibitem{Balitsky:1996ub}
I.~Balitsky, {\it Operator expansion for high-energy scattering},  {\em Nucl.
  Phys.} {\bf B463} (1996) 99--160,
  [\href{http://xxx.lanl.gov/abs/hep-ph/9509348}{{\tt hep-ph/9509348}}].

\bibitem{Kovchegov:1997pc}
Y.~V. Kovchegov, {\it Quantum structure of the non-abelian
  {Weizsaecker-Williams} field for a very large nucleus},  {\em Phys. Rev.}
  {\bf D55} (1997) 5445--5455,
  [\href{http://xxx.lanl.gov/abs/hep-ph/9701229}{{\tt hep-ph/9701229}}].

\bibitem{Jalilian-Marian:1997xn}
J.~Jalilian-Marian, A.~Kovner, L.~D. McLerran, and H.~Weigert, {\it The
  intrinsic glue distribution at very small x},  {\em Phys. Rev.} {\bf D55}
  (1997) 5414--5428, [\href{http://xxx.lanl.gov/abs/hep-ph/9606337}{{\tt
  hep-ph/9606337}}].

\bibitem{Jalilian-Marian:1997jx}
J.~Jalilian-Marian, A.~Kovner, A.~Leonidov, and H.~Weigert, {\it The {BFKL}
  equation from the {Wilson} renormalization group},  {\em Nucl. Phys.} {\bf
  B504} (1997) 415--431, [\href{http://xxx.lanl.gov/abs/hep-ph/9701284}{{\tt
  hep-ph/9701284}}].

\bibitem{Jalilian-Marian:1997gr}
J.~Jalilian-Marian, A.~Kovner, A.~Leonidov, and H.~Weigert, {\it The {Wilson}
  renormalization group for low x physics: Towards the high density regime},
  {\em Phys. Rev.} {\bf D59} (1999) 014014,
  [\href{http://xxx.lanl.gov/abs/hep-ph/9706377}{{\tt hep-ph/9706377}}].

\bibitem{Jalilian-Marian:1997dw}
J.~Jalilian-Marian, A.~Kovner, and H.~Weigert, {\it The {Wilson}
  renormalization group for low x physics: Gluon evolution at finite parton
  density},  {\em Phys. Rev.} {\bf D59} (1999) 014015,
  [\href{http://xxx.lanl.gov/abs/hep-ph/9709432}{{\tt hep-ph/9709432}}].

\bibitem{Balitsky:1997mk}
I.~Balitsky, {\it Operator expansion for diffractive high-energy scattering},
  \href{http://xxx.lanl.gov/abs/hep-ph/9706411}{{\tt hep-ph/9706411}}.

\bibitem{Jalilian-Marian:1998cb}
J.~Jalilian-Marian, A.~Kovner, A.~Leonidov, and H.~Weigert, {\it Unitarization
  of gluon distribution in the doubly logarithmic regime at high density},
  {\em Phys. Rev.} {\bf D59} (1999) 034007,
  [\href{http://xxx.lanl.gov/abs/hep-ph/9807462}{{\tt hep-ph/9807462}}].

\bibitem{Mueller:1999wm}
A.~H. Mueller, {\it Parton saturation at small x and in large nuclei},  {\em
  Nucl. Phys.} {\bf B558} (1999) 285--303,
  [\href{http://xxx.lanl.gov/abs/hep-ph/9904404}{{\tt hep-ph/9904404}}].

\bibitem{Kovchegov:1999ua}
Y.~V. Kovchegov, {\it Unitarization of the {BFKL} pomeron on a nucleus},  {\em
  Phys. Rev.} {\bf D61} (2000) 074018,
  [\href{http://xxx.lanl.gov/abs/hep-ph/9905214}{{\tt hep-ph/9905214}}].

\bibitem{Kovner:2000pt}
A.~Kovner, J.~G. Milhano, and H.~Weigert, {\it Relating different approaches to
  nonlinear {QCD} evolution at finite gluon density},  {\em Phys. Rev.} {\bf
  D62} (2000) 114005, [\href{http://xxx.lanl.gov/abs/hep-ph/0004014}{{\tt
  hep-ph/0004014}}].

\bibitem{Weigert:2000gi}
H.~Weigert, {\it Unitarity at small {B}jorken x},  {\em Nucl. Phys.} {\bf A703}
  (2002) 823--860, [\href{http://xxx.lanl.gov/abs/hep-ph/0004044}{{\tt
  hep-ph/0004044}}].

\bibitem{Iancu:2000hn}
E.~Iancu, A.~Leonidov, and L.~D. McLerran, {\it Nonlinear gluon evolution in
  the color glass condensate. {I}},  {\em Nucl. Phys.} {\bf A692} (2001)
  583--645, [\href{http://xxx.lanl.gov/abs/hep-ph/0011241}{{\tt
  hep-ph/0011241}}].

\bibitem{Ferreiro:2001qy}
E.~Ferreiro, E.~Iancu, A.~Leonidov, and L.~McLerran, {\it Nonlinear gluon
  evolution in the color glass condensate. {II}},  {\em Nucl. Phys.} {\bf A703}
  (2002) 489--538, [\href{http://xxx.lanl.gov/abs/hep-ph/0109115}{{\tt
  hep-ph/0109115}}].

\bibitem{Golec-Biernat:1998js}
K.~Golec-Biernat and M.~{W\"usthoff}, {\it Saturation effects in deep inelastic
  scattering at low {$Q^2$} and its implications on diffraction},  {\em Phys.
  Rev.} {\bf D59} (1999) 014017,
  [\href{http://xxx.lanl.gov/abs/hep-ph/9807513}{{\tt hep-ph/9807513}}].

\bibitem{Golec-Biernat:1999qd}
K.~Golec-Biernat and M.~{W\"usthoff}, {\it Saturation in diffractive deep
  inelastic scattering},  {\em Phys. Rev.} {\bf D60} (1999) 114023,
  [\href{http://xxx.lanl.gov/abs/hep-ph/9903358}{{\tt hep-ph/9903358}}].

\bibitem{Stasto:2000er}
A.~M. Stasto, K.~Golec-Biernat, and J.~Kwiecinski, {\it Geometric scaling for
  the total $\gamma^* p$ cross-section in the low x region},  {\em Phys. Rev.
  Lett.} {\bf 86} (2001) 596--599,
  [\href{http://xxx.lanl.gov/abs/hep-ph/0007192}{{\tt hep-ph/0007192}}].

\bibitem{Blaizot:2002xy}
J.-P. Blaizot, E.~Iancu, and H.~Weigert, {\it Non linear gluon evolution in
  path-integral form},  \href{http://xxx.lanl.gov/abs/hep-ph/0206279}{{\tt
  hep-ph/0206279}}.

\bibitem{Golec-Biernat:2001if}
K.~Golec-Biernat, L.~Motyka, and A.~M. Stasto, {\it Diffusion into infra-red
  and unitarization of the {BFKL} pomeron},
  \href{http://xxx.lanl.gov/abs/hep-ph/0110325}{{\tt hep-ph/0110325}}.

\bibitem{Braun:2000wr}
M.~Braun, {\it Structure function of the nucleus in the perturbative {QCD} with
  ${N}_c \to \infty$ ({BFKL} pomeron fan diagrams)},  {\em Eur. Phys. J.} {\bf
  C16} (2000) 337--347, [\href{http://xxx.lanl.gov/abs/hep-ph/0001268}{{\tt
  hep-ph/0001268}}].

\bibitem{Iancu:2002tr}
E.~Iancu, K.~Itakura, and L.~McLerran, {\it Geometric scaling above the
  saturation scale},  {\em Nucl. Phys.} {\bf A708} (2002) 327--352,
  [\href{http://xxx.lanl.gov/abs/hep-ph/0203137}{{\tt hep-ph/0203137}}].

\bibitem{Braun:2003un}
M.~A. Braun, {\it Pomeron fan diagrams with an infrared cutoff and running
  coupling},  \href{http://xxx.lanl.gov/abs/hep-ph/0308320}{{\tt
  hep-ph/0308320}}.

\bibitem{Gotsman:2002yy}
E.~Gotsman, E.~Levin, M.~Lublinsky, and U.~Maor, {\it Towards a new global
  {QCD} analysis: Low x {DIS} data from non- linear evolution},  {\em Eur.
  Phys. J.} {\bf C27} (2003) 411--425,
  [\href{http://xxx.lanl.gov/abs/hep-ph/0209074}{{\tt hep-ph/0209074}}].

\bibitem{Fadin:1995xg}
V.~S. Fadin, M.~I. Kotsky, and R.~Fiore, {\it Gluon reggeization in qcd in the
  next-to-leading order},  {\em Phys. Lett.} {\bf B359} (1995) 181--188.

\bibitem{Fadin:1996zv}
V.~S. Fadin, M.~I. Kotsky, and L.~N. Lipatov, {\it Gluon pair production in the
  quasi-multi-regge kinematics},
  \href{http://xxx.lanl.gov/abs/hep-ph/9704267}{{\tt hep-ph/9704267}}.

\bibitem{Fadin:1998hr}
V.~S. Fadin, R.~Fiore, A.~Flachi, and M.~I. Kotsky, {\it Quark-antiquark
  contribution to the {BFKL} kernel},  {\em Phys. Lett.} {\bf B422} (1998)
  287--293, [\href{http://xxx.lanl.gov/abs/hep-ph/9711427}{{\tt
  hep-ph/9711427}}].

\bibitem{Fadin:1998py}
V.~S. Fadin and L.~N. Lipatov, {\it {BFKL} pomeron in the next-to-leading
  approximation},  {\em Phys. Lett.} {\bf B429} (1998) 127--134,
  [\href{http://xxx.lanl.gov/abs/hep-ph/9802290}{{\tt hep-ph/9802290}}].

\bibitem{Colferai:1999em}
D.~Colferai, {\it Small-x processes in perturbative quantum chromodynamics},
  \href{http://xxx.lanl.gov/abs/hep-ph/0008309}{{\tt hep-ph/0008309}}.

\bibitem{Thorne:1999rb}
R.~S. Thorne, {\it {NLO} {BFKL} equation, running coupling and renormalization
  scales},  {\em Phys. Rev.} {\bf D60} (1999) 054031,
  [\href{http://xxx.lanl.gov/abs/hep-ph/9901331}{{\tt hep-ph/9901331}}].

\bibitem{Thorne:2001nr}
R.~S. Thorne, {\it The running coupling {BFKL} anomalous dimensions and
  splitting functions},  {\em Phys. Rev.} {\bf D64} (2001) 074005,
  [\href{http://xxx.lanl.gov/abs/hep-ph/0103210}{{\tt hep-ph/0103210}}].

\bibitem{Mueller:2002zm}
A.~H. Mueller and D.~N. Triantafyllopoulos, {\it The energy dependence of the
  saturation momentum},  {\em Nucl. Phys.} {\bf B640} (2002) 331--350,
  [\href{http://xxx.lanl.gov/abs/hep-ph/0205167}{{\tt hep-ph/0205167}}].

\bibitem{Triantafyllopoulos:2002nz}
D.~N. Triantafyllopoulos, {\it The energy dependence of the saturation momentum
  from {RG} improved {BFKL} evolution},  {\em Nucl. Phys.} {\bf B648} (2003)
  293--316, [\href{http://xxx.lanl.gov/abs/hep-ph/0209121}{{\tt
  hep-ph/0209121}}].

\bibitem{Freund:2002ux}
A.~Freund, K.~Rummukainen, H.~Weigert, and A.~{Sch\"afer}, {\it Geometric
  scaling in inclusive {$e A$} reactions and nonlinear perturbative {QCD}},
  {\em Phys. Rev. Lett.} {\bf 90} (2003) 222002,
  [\href{http://xxx.lanl.gov/abs/hep-ph/0210139}{{\tt hep-ph/0210139}}].

\bibitem{Mueller:2003bz}
A.~H. Mueller, {\it Nuclear {A}-dependence near the saturation boundary},
  \href{http://xxx.lanl.gov/abs/hep-ph/0301109}{{\tt hep-ph/0301109}}.

\end{thebibliography}

\providecommand{\href}[2]{#2}\begingroup\raggedright\endgroup

\end{document}